\documentclass[12pt,preprint]{aastex}

\begin{document}

\title{
Microlensing optical depth towards the Galactic bulge from MOA
observations during 2000 with Difference Image Analysis }

\author
{
T. Sumi\altaffilmark{1,2},
F. Abe\altaffilmark{2},
I.A. Bond\altaffilmark{3},
R.J. Dodd\altaffilmark{4,12},
J.B. Hearnshaw\altaffilmark{5},
 M.Honda \altaffilmark{6},
M.Honma\altaffilmark{7},
Y.Kan-ya\altaffilmark{7},
P.M.Kilmartin\altaffilmark{5},
K.Masuda\altaffilmark{2},
Y.Matsubara\altaffilmark{2},
Y.Muraki\altaffilmark{2},
T.Nakamura\altaffilmark{8},
R. Nishi\altaffilmark{9},
S. Noda\altaffilmark{7},
K.Ohnishi\altaffilmark{10},
O.K.L. Petterson\altaffilmark{5}
N.J.Rattenbury\altaffilmark{11},
M.Reid\altaffilmark{12},
To.Saito\altaffilmark{13},
Y. Saito\altaffilmark{2},
H.Sato\altaffilmark{14},
M.Sekiguchi\altaffilmark{6},
J.Skuljan\altaffilmark{5},
D.J.Sullivan\altaffilmark{12},
M.Takeuti\altaffilmark{15},
P.J. Tristram\altaffilmark{5},
S. Wilkinson\altaffilmark{12},
T.Yanagisawa\altaffilmark{16},
P.C.M.Yock\altaffilmark{11}
}
\altaffiltext{1}{Princeton University Observatory, Princeton, NJ 08544-1001, USA}
\altaffiltext{2}{Solar-Terrestrial Environment Laboratory,  Nagoya University,
Nagoya 464-8601, Japan }
\altaffiltext{3}{Inst. of Astronomy, University of Edinburgh, Royal Observatory, Edinburgh, UK}
\altaffiltext{4}{Carter National Observatory, Wellington, NZ}
\altaffiltext{5}{Dept. of Physics and Astronomy, University of Canterbury, Christchurch, NZ}
\altaffiltext{6}{Inst. Cosmic Ray Research, University of Tokyo, chiba 277-0882, Japan}
\altaffiltext{7}{National Astronomical Observatory, 2-21-1 Osawa, Mitaka, Tokyo 181-8588, Japan}
\altaffiltext{8}{Dept. of Physics, Kyoto University, Koto 606-8502, Japan}
\altaffiltext{9}{Dept. of Physics, Niigata University, Niigata 950-2181, Japan}
\altaffiltext{10}{Nagano National College of Technology, Japan}
\altaffiltext{11}{Dept. of Physics, University of Auckland, Auckland, NZ}
\altaffiltext{12}{School of Chemical and Physical Sciences, Victoria University, Wellington, NZ}
\altaffiltext{13}{Tokyo Metropolitan College of Aeronautics, Tokyo 140-0011, Japan}
\altaffiltext{14}{Dept. of Physics, Kounan University, Koube 658-8501, Japan}
\altaffiltext{15}{Astronomical Institute, Tohoku University, Sendai 980-8578, Japan}
\altaffiltext{16}{National Aerospace Laboratory of Japan, Tokyo 182-8522, Japan}

\begin{abstract}
We analyze the data of the gravitational microlensing survey
carried out by  by the MOA group during 2000
towards the Galactic Bulge (GB).
Our observations are designed to detect efficiently 
high magnification events  with faint source stars and
short timescale events, by increasing the the sampling rate up to
$\sim 6$ times per night and using Difference Image Analysis
(DIA). We detect $28$ microlensing candidates in
$12$ GB fields corresponding to $16$ deg$^2$. We use Monte
Carlo simulations to estimate our microlensing event detection
efficiency, where we construct the $I$-band extinction map of our GB
fields in order to find dereddened magnitudes.
We find a systematic bias and large uncertainty in the measured value
of the timescale $t_{\rm Eout}$ in our simulations. They are 
associated with blending and unresolved sources, and are allowed 
for in our measurements.  We compute an optical depth 
$\tau = 2.59_{-0.64}^{+0.84} \times 10^{-6}$ towards the GB for events 
with timescales $0.3<t_{\rm E}<200$ days.  We consider disk-disk lensing, 
and obtain an optical depth  
$\tau_{bulge} =  3.36_{-0.81}^{+1.11}\times 10^{-6}[0.77/(1-f_{\rm disk})]$
for the bulge component
assuming a $23\%$ stellar contribution from disk stars. These
observed optical depths are consistent with previous
measurements by the MACHO and OGLE groups, and still higher than those
predicted by existing Galactic models. We present the timescale
distribution of the observed events, and find there are no
significant short events of a few days, in spite of
our high detection efficiency for short timescale events down to
$t_{\rm E} \sim 0.3$ days. We find that half of all our
 detected events have high magnification ($>10$). These events are
useful for studies of extra-solar planets.

\end{abstract}


\section{Introduction}
\label{sec:intro} 
Following the suggestion of \cite{pac91} and \cite{gri91},
several groups have carried out microlensing
surveys towards the Galactic Bulge (GB), as seen in Baade's window.
It is now well understood that these observations are useful for
studying the structure, dynamics and kinematics of the Galaxy and
the stellar mass function as the event rate and timescale
distributions are related to the masses and velocities of lens
objects.

The amplification of a microlensing event is described by (\citealt{pac86})
\begin{equation}
  \label{eq:amp-u}
  A(u)= \frac{u^2+2}{u\sqrt{u^2+4}},
\end{equation}
where $u$ is the projected separation of the source and lens in units of
the Einstein radius $R_{\rm E}$ which is given by
\begin{equation}
  R_{\rm E}(M,x) = \sqrt{\frac{4GM}{c^2}D_{\rm s}x(1-x)},
  \label{eq:re}
\end{equation}
where $M$ is the lens mass, $x=D_{\rm l}/D_{\rm s}$ is the normalized
lens distance and $D_{\rm l}$ and $D_{\rm s}$ are the observer-lens and
the observer-source star distances.
The time variation of $u=u(t)$ is
\begin{equation}
  \label{eq:u}
  u(t)=\sqrt{\beta^{2} + \left( \frac{t-t_{0}}{t_{\rm E}} \right)^2},
\end{equation}
where $\beta$, $t_{0}$, $t_{\rm E}= R_{\rm E}/v_{\rm t}$ and
$v_{\rm t}$ are the minimum impact parameter in units of $R_{\rm E}$,
the time of maximum magnification, the event time scale and the
transverse velocity of the lens relative to the line of sight
towards the source star, respectively. From a light curve, one can
determine the values of $\beta$, $t_{0}$ and $t_{\rm E}$, but not 
the values of $M$, $x$ or $v_{\rm t}$.

Our MOA (Microlensing Observations in Astrophysics)
group started observations towards the GB in 1999.
From 2000 we introduced the Difference Image Analysis (DIA) (\citealt{cro92};
\citealt{phi95}; \citealt{ala98}; \citealt{ala00}; \citealt{alc99,alc00};
\citealt{woz00}; \citealt{bon01})
which is able to perform better photometry than the traditional
DoPHOT (\citealt{sch93}) type analysis in crowded fields
at any place even where no star was identified.

To date, hundreds of microlensing events have been detected towards
the GB by the OGLE (\citealt{uda94,uda00}; \citealt{woz01}) and MACHO 
collaborations (\citealt{alc97a,alc00}). They estimate the microlensing 
optical depth towards the GB to be $3.3 \pm1.2 \times 10^{-6}$ from $9$ 
events by DoPHOT analysis,
$3.9 ^{+1.8}_{-1.2} \times 10^{-6}$ from $13$ events in a clump giant
subsample from DoPHOT, and  $3.23^{+0.52}_{-0.50} \times 10^{-6}$ from 
$99$ events by DIA respectively. \cite{pop00} and  \cite{pop02}
estimate values of $2.0 ^{+0.4}_{-0.4} \times 10^{-6}$ and  
$2.23 ^{+0.38}_{-0.35} \times 10^{-6}$ respectively from MACHO data.
These values are all more than twice those expected from existing 
Galactic models, which are somewhere around $0.5 \sim  1.0 \times 10^{-6}$
(\citealt{pac91}; \citealt{gri91}; \citealt{kir94}).  This suggests
that the standard models of the Galaxy need to be revised.
To explain the high optical depth, a number of authors have suggested
the presence of a bar oriented along our line of sight to the GB
(\citealt{pac94}; \citealt{zha95}), and have adopted various values of the bar
orientation and mass (\citealt{pac94}; \citealt{pea98}; \citealt{zha96}).
Microlensing observations towards the GB therefore appear useful for
characterizing the mass and inclination of the bar.

\cite{pop00} and  \cite{pop02} raised the possibility
of a systematic bias in the optical depth due to the difficulties
of measuring $t_{\rm E}$ associated with blending and unresolved
sources.  When the actual source base-line flux is unknown,
$t_{\rm E}$ and $\beta$ are degenerate in relatively low signal-to-noise 
ratio (S/N) events (c.f. \citealt{han99}; \citealt{bon01};\citealt{gou02}).
The optical depth may be estimated by using red clump giant stars 
to avoid the bias, or by other methods (e.g. \citealt{gon99}; \citealt{ker01}).
In this paper, we quantify the bias by using Monte Carlo simulations and take itinto account to estimate the optical depth.

Our observations are designed to detect efficiently high
magnification events with faint source stars for the study of
extra-solar planets and surface-transit events, by increasing 
the sampling rate up to $\sim 6$ times per night (note that the 
sampling rate of other projects is typically once per night).
Our observations are consequently fairly sensitive to short duration
events, i.e., events caused by smaller mass lenses. This could lead to 
a different optical depth estimate from previous studies if there is a 
significant contribution of low mass objects such as brown dwarfs to the
microlensing optical depth. Thus the MOA observations  
can constrain the contribution of the low-mass population and also the
structure of the GB. For this purpose
here we present the results of MOA observations.

In this paper, the results of the DIA analysis of data towards the GB
taken by MOA in 2000 with DIA are presented. The analysis is
aimed at finding how efficiently we can detect high
magnification events and short timescale events, and at estimating
the optical depth towards the GB.
In \S\,\ref{sec:observation} we describe
our observations. \S\,\ref{sec:dataanalysis} is devoted to
the analysis method. In \S\,\ref{sec:results} we describe the
event selection process and results. In \S\,\ref{sec:extinction}
we make an $I$-band extinction map of our GB fields caused by
dust. This is useful in estimating extinction free source
magnitudes in the simulation. In \S\,\ref{sec:OPTICALDEPTH} we
describe the simulation used to estimate our detection efficiency
and the resultant microlensing optical depth. Discussion and
conclusions are given in \S\,\ref{sec:disc}.

\section{Observations}
\label{sec:observation}

We observed from the Mt.John University Observatory (MJUO) in New Zealand
at $170^{\circ} 27.9'$E, $43^{\circ}59.2'$S, $1030$ m altitude.
Typical sky background values are $21.9, 22.6, 21.5, 20.9$ and $19.1$
mag\,arcsec$^{-2}$ in the U, B, V, Rc and Ic passbands
respectively at an air mass of $1.0$ (Gilmore, private communication, 1994).
Spectroscopic hours are $\sim45\%$. 

We used a 61-cm telescope equipped with  the large mosaic CCD
camera MOA-cam2 (\citealt{yan00}). This  has three 2k $\times$ 4k
pixels thinned CCDs. The combined field of view of the camera and
telescope is $0.92^\circ \times 1.39^\circ$ ($0.81$
arcsec\,pixel$^{-1}$). We used two non-standard wide passband Red
($630-1100$ nm) and Blue ($400-630$ nm) filters.
The global seeing is typically $1.9-3.5''$.

We observed 14 GB fields ($18$ deg$^2$) $3 \sim 6$ times per night 
during the southern winter season (from April to November),
where each field consists of three subfields corresponding to each
CCD. The main aim of the observations was to detect high
magnification events in order to find extrasolar planets
(\citealt{bon01}). The detection probability of extrasolar planets
in high magnification events is high (\citealt{gri98}).

Microlensing events in which the source star is fainter than the
observational limiting magnitude, are generally known as ``Pixel Lensing''
events (\citealt{gou96}) or ``EAGLE events'' (\citealt{nak98};
\citealt{sum00}). These usually have high magnification. However, the
period in which they are visible is very short,
because they are visible only at peak of magnification. 
To detect this kind of events we sample each
field $5 \sim 6$ times per day mainly through the Red filter
rather than both colors. Due to this high sampling rate, the
detection efficiency for short timescale events also becomes
higher than the former studies. A unique feature of our
observations is the ability to detect small lens objects towards
the GB.

\section{DATA Analysis}
\label{sec:dataanalysis}

We have used the DIA technique to provide sensitivity to Pixel Lensing
or EAGLE events. These events cannot be detected by the conventional 
(fixed position) DoPHOT-type photometry.

We developed our own implementation of the DIA method of
\cite{ala98} and \cite{ala00} which directly model the kernel 
in real space with spatial variations of the kernel across the CCD.
This method is suitable even for crowded and poor S/N images.
In this technique, one first registers some star positions, and
geometrically aligns each ``current'' observation image to a
preselected ``reference'' image which is recorded in good seeing,
high S/N and low airmass. Next, the
convolution kernel which is required to map the reference image to
the current image, is calculated by using the current and
reference images. The reference images are convolved to match the
seeing and scaled to align with the current images. The resultant
images are then subtracted and variable objects are
detected.

The images were analyzed in real-time at MJUO to issue alerts
for the events (\citealt{bon01}).  In this analysis we use the same 
sample of subtracted images as in the real-time analysis.

We have found the effect of differential refraction, which causes a
significant residual flux in the subtracted image
(\citealt{tom96}; \citealt{alc99}), is negligible in our images
with the red filter when the airmass is lower than $2.0$. We
have not corrected for this effect because our data are mainly taken
through the red filter.

We have made a catalogue of all stars in the reference images by using
DoPHOT. All saturated stars and bad pixels on the
reference images are masked out the same as in the subtraction
process. These are useful in the following analyses.

\subsection{Identification of variable objects}
\label{sec:detect}

On a subtracted image, variable objects can be seen as
positive or negative profiles depending on whether
the flux had increased or decreased relative to the reference
image.

To detect these objects, we use our implementation of the
algorithm in the IRAF task DAOFIND, in which both positive and
negative profiles could be detected simultaneously. There are also
spurious profiles not associated with stellar variability, for
example, cosmic rays, satellite tracks and electrons leaked from
bright saturated star images. To avoid detecting these spurious
objects, we apply several criteria about the statistics and
profiles in the Analog Digital Unit (ADU) value of pixels around
the peak, e.g., S/N is required to be larger than 3.

Candidate objects which pass these criteria are checked
against those obtained in previous reductions of the field.
If no object is cross-referenced, these new objects are added to
the initial list of variable object positions. If an object has previously been
detected, this is identified as the same object and the number of detections
for this object is incremented. The number of detections for each
object is used in the event selection cut ``cut1'' (see \S\,\ref{sec:cut1}).

\subsection{Photometry}
\label{sec:photometry}

To the list of variable objects made in the previous section, we
have applied the first simple event selection  ``cut1'' (see
\S\,\ref{sec:cut1}) to cull the number of spurious objects in this
list.

For the objects which passed the ``cut1'' test, we have performed PSF
profile fitting photometry, where the high S/N empirical PSF
images ($23 \times 23$ pixels) is made for each of the 32
($500\times 500$ pixels) sub-regions in the reference image. Then
these PSF images are convolved by the same kernel function used in
the subtraction process, to match this PSF to that in each local
position of each time series of frames. And the total flux of this
PSF is normalized to $1$.

The PSF photometry comes down to a two-parameter fit for the
amplitude $a_{psf}$ and base line $b$, i.e. $a_{psf}PSF_i + b$,
here $PSF_i$ is the value of empirical PSF image in pixel $i$
centered on the variable object. 
In principal, in subtracted images, the background is zero by 
virtue of the least-square process. However, in fact, there are 
residuals especially around the very bright saturated stars, 
bright variable stars and in poor S/N images.  In our images, 
the fraction of the area where these effects occur is small but not 
negligible because of the wide wing of the PSF (our median 
seeing is about $2''.3$).  When we compared the 1-parameter and 
2-parameter fits, we found that several faint stars around a 
bright variable star show similar variability as the bright variable 
star in the 1-parameter fit, but not in the 2-parameter fit.  
Fitting a 2nd "base line flux" parameter on the subtracted
images provides a good zero-point check.
So we make the base line $b$ a fitting parameter.
In the subtracted images the flux
has been rescaled to match that in the reference image by dividing
with the scale factor $a_0$ which is the first coefficient of the
kernel function of the frame. By this process, all the time series
of subtracted images are photometrically calibrated. Then no
extinction correction is required. In the fitting, the noise of
each pixel $\sigma_i$ of the subtracted image is given from the
actual flux $f_{i}$ in the current image before subtraction,
taking account of the gain $G$ in ADU/e$^-$ and the scale factor
$a_0$ of each frame.

This photometry has been performed on all objects in the variable list
in all the time series of images and the differential flux $\Delta
F = a_{psf}$ light curves stored in the database with their
corresponding error and the square root of the reduced $\chi^2$ in
the PSF fitting (we refer to this as $Sdev_{psf}$).

\subsubsection{Noise properties}
\label{sec:noise}

The error in the PSF photometry $\sigma_{psf}$ obtained by $\chi^2$ fitting
would be optimistic, as it includes only the photon noise
component.  So we test the properties of the photometry using a sample
of constant stars.
We have randomly sampled $\sim 1000$ constant stars for each image from our
star catalogue, in which we have rejected the variable objects detected in
\S\,\ref{sec:detect}.

Noise properties are derived from the residuals of individual measurements
around the mean flux of each object. So we use $\sim 100,000$ 
measurements for each chip.  Each residual is normalized by the 
error from the photon noise $\sigma_{psf}$
for the corresponding photometric point. Then stars are grouped
according to their brightness on the reference image,
the standard deviation normalized by $\sigma_{psf}$ in the PSF fitting, i.e.
$Sdev_{psf}$.
We think that the $Sdev_{psf}$ of each measurement should be a good indicator
for the systematic noise which comes from the non-photon noise.
All residuals coming from the light curves of stars in a given group
are merged into one distribution.

\begin{figure}
\epsscale{0.8}
\includegraphics[angle=-90, scale=0.65]{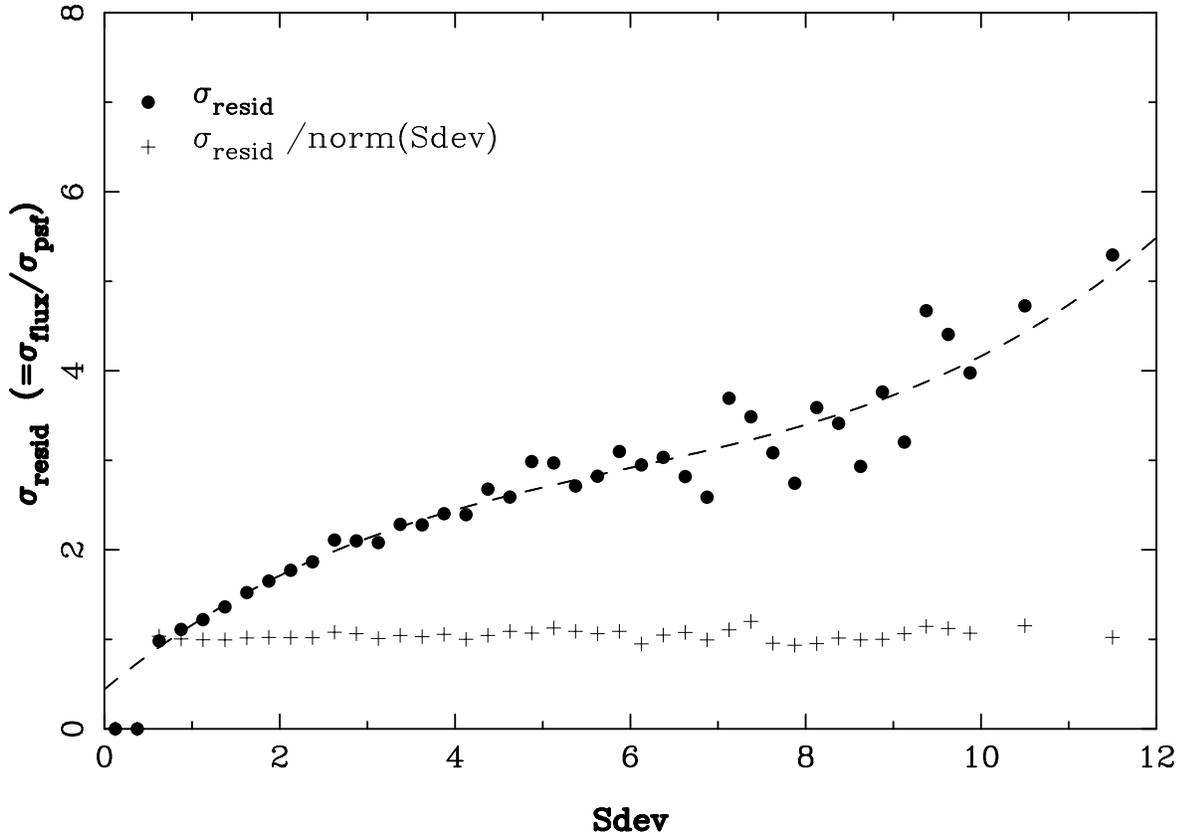}
\caption{
Half width of the region containing $68\%$ of residuals
which are a ratio of the actual scatter to the photon
noise estimate as a function of  $Sdev_{psf}$ (filled
circle) and the same plot with renormalization (cross).
The dashed line indicates the best fit for the residual
plot with 3-degree polynomial.
\label{fig:sdev-sig} }
\end{figure}

For each group we calculate the half width of the region
containing $68.3\%$ of residuals $\sigma_{resid}$, a robust
estimator of the width $\sigma$ of the Gaussian distribution. In
Figure \ref{fig:sdev-sig}, we plot the estimated $\sigma_{resid}$
as a function of $Sdev_{psf}$ and these are fitted  with a
3-degree polynomial. Here bright stars ($R < -11$ in DoPHOT
output red magnitude, which corresponds to $I < 15$; see 
\S\,\ref{sec:calibration}) are rejected
because such bright stars have other systematic deviations as
discussed below. The effect of such bright stars is small since,
in total, they represent less than $0.4\%$ of all possible source
stars ($I<23$), and the pixels where such stars dominate on the
CCD chip is less than $1\%$. We did not use those measurements 
with $Sdev_{psf} >10$ as they are unreliable. The fraction of such 
measurements is $\sim 7\%$ out of all measurements. In
this figure, a clear trend for these values can be seen, and the
fitted curve  can be used to rescale the error bars in order to
improve the consistency of the light curves. We also show the same
plot normalized with this fitted function in Figure
\ref{fig:sdev-sig}.

\begin{figure}
\plotone{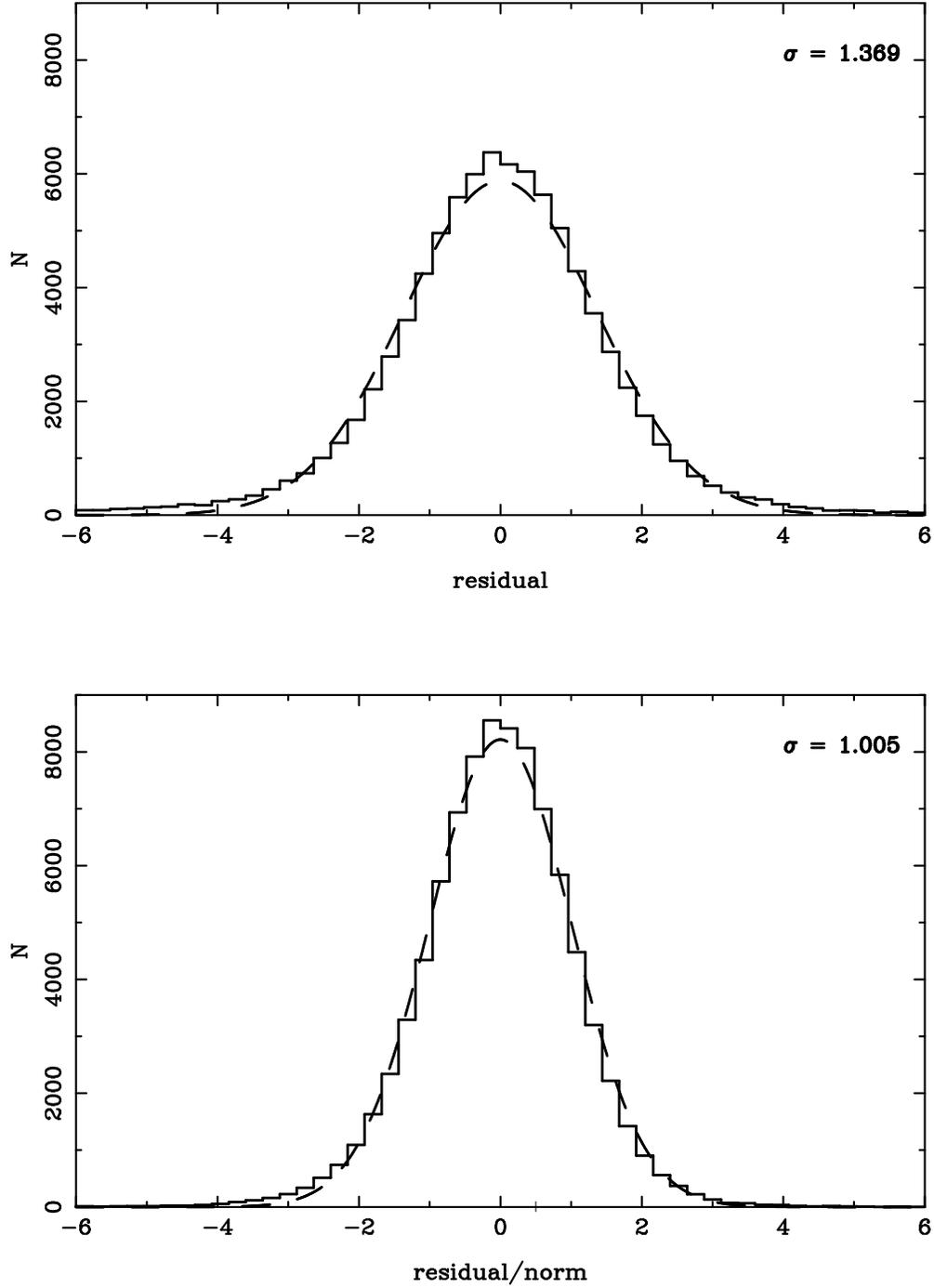}
\caption{Distribution of residuals (upper) and renormalized
residuals (lower) for 90,000 individual measurements of constant stars
as histograms.  The dotted lines indicate the best fit Gaussian
distributions centered on 0.  The half widths of the regions containing
$68\%$ of the residuals $\sigma$ are $1.369$ and $1.005$ respectively.}
\label{fig:residual}
\end{figure}

To check whether this normalization is appropriate or not, we show
the distribution histograms of these residuals (upper panel) and
normalized residuals (lower panel) in Figure \ref{fig:residual}.
As shown in Figure \ref{fig:residual}, the distribution of
normalized residuals is consistent with the normal distribution.
So our normalization seems to be reasonable. This exercise have been 
carried out for each of the CCD chips which showed that chips 2
and 3 are very similar but differ slightly from chip 1. So we have
estimated this function for each chip. This trend is very similar
over all fields measured with the same CCD chip. We have applied this
normalization to all our measurements.

\begin{figure}
\includegraphics[angle=-90, scale=0.65]{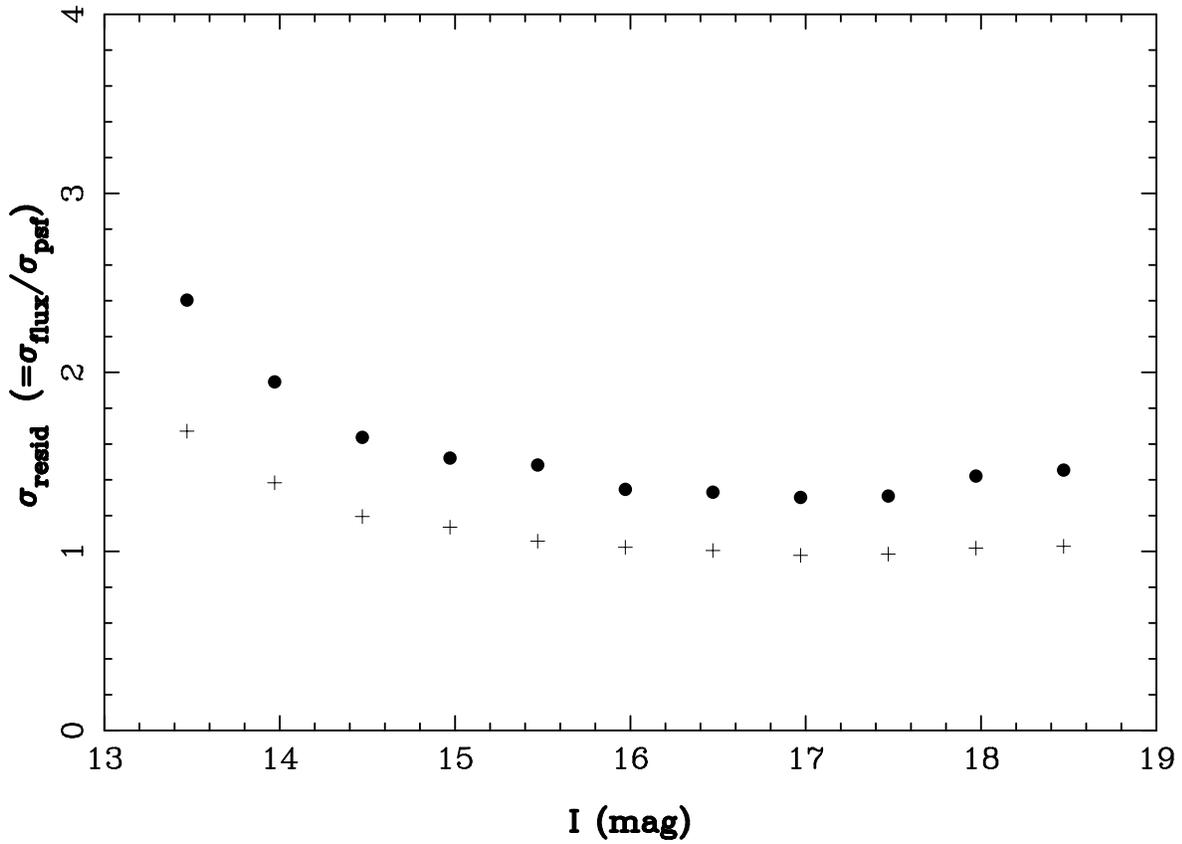}
\caption{ Half width of the region containing $68\%$ of the
residuals which are a ratio of the actual scatter to the error
from the photon noise (filled circle) as a function of  $I$-band
magnitude and the same plot with
renormalization (cross). The data in this plot come from $90,000$
individual measurements of $1,000$ constant stars.
\label{fig:mag-sig} }
\end{figure}

We also show the $\sigma_{resid}$ of residuals (filled circle) and
normalized residuals (cross) as a function of brightness of
constant stars in $I$-band magnitude (see \S\,\ref{sec:calibration}) 
in Figure \ref{fig:mag-sig}. As
shown in Figure \ref{fig:mag-sig}, as we move towards brighter
stars the $\sigma_{resid}$ increases due to systematic effects
related to the uncertainty in the seeing and the PSF, first slowly
and then faster. \cite{ala98} provide a possible explanation in
terms of the atmospheric turbulence. Another possible source of
this excess is the existence of some small amplitude variable
stars in these bright stars. No correction for this effect is 
made in this analysis. In Figure \ref{fig:mag-sig}, we find that
the noise (standard deviation) in this analysis is about $40\%$
above the photon noise limit.

\subsection{Calibration of Fluxes}
\label{sec:calibration}

To obtain the source magnitude from the baseline flux $f_0$ given
by light curve fitting, we need the transformation relation
between the flux in the reference image and the apparent magnitude
in standard passbands. We use the UBVI photometry catalogue of
selected stars in Baade's window provided by OGLE
(\citealt{pac99}) that are contained in three MOA GB subfields
(ngb2-1, ngb2-2, ngb3-3, where ngb2-1 means chip 1 in field ngb2).
In these regions, the extinction is relatively low and uniform.
Our star catalogue have been made by applying DoPHOT to half of all the
frames for each field which contain $\sim 20$ Blue images taken
during the 2000 season. Then we have made the $R (MOA)$ and $B (MOA)$
catalogue of all stars in our GB fields by taking the median for
each star so as to increase the accuracy of the photometry and to
avoid the daily difference of extinction by the atmosphere. Then
the DoPHOT photometry $R_{ref}$ of the reference images is 
compared with the catalogue value $R_{med}$. The differences are
$R_{ref}- R_{med}= -0.1 \sim +0.03$ mag and spatial variations in
one chip are about $\sim 0.02$ mag. So we have neglected local
differences in the chip and stored one offset value for each chip.
These offsets are used in the following calibrations.

\begin{figure}
\includegraphics[angle=-90, scale=0.7]{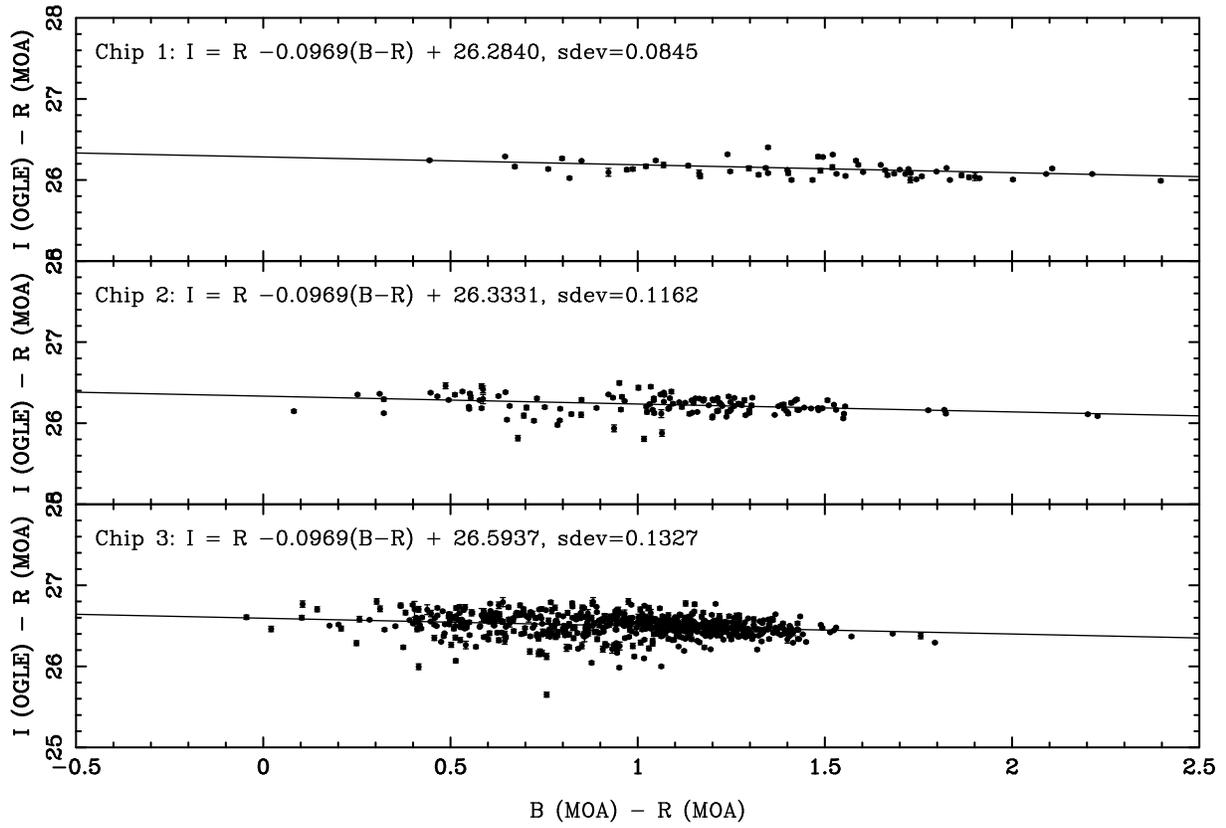}
\caption{Calibration of MOA Red measurements to the standard $I$ band
magnitude for the three CCDs (chip1, 2 and 3 from top to bottom)
with the color term.
\label{fig:Rmed-I_color}}
\end{figure}

\begin{figure}
\includegraphics[angle=-90, scale=0.7]{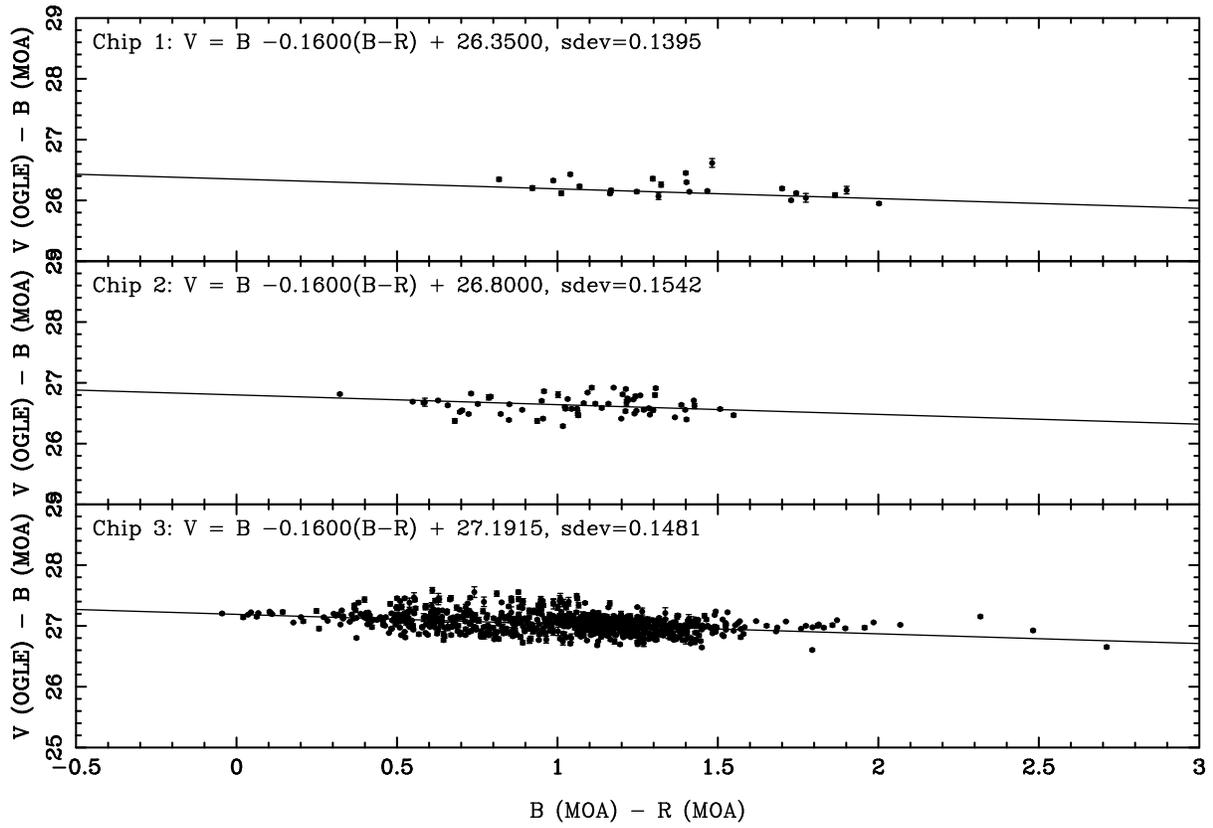}
\caption{Calibration of MOA Blue measurements to the standard $V$ band
magnitude scale of the three CCDs of the camera MOA-cam2 with the
color term.
\label{fig:Bmed-V_color}}
\end{figure}

Next, the $R$ and $B$ measurements in our catalogue of $3$ MOA-OGLE
overlap fields are compared with the $V$ and $I$ photometry of the 
corresponding star in the OGLE catalogue.
Fundamentally the transformation to the standard $V$ and $I$ from MOA
non-standard $R$ and $B$ depends on the color index $B - R$.
The transformations obtained to the standard $I$ and $V$ for each chip
are shown in Figure \ref{fig:Rmed-I_color} and \ref{fig:Bmed-V_color},
and are given by following equations.

\begin{eqnarray}
\label{eq:R2Icolor}
I = R - 0.0969 (B - R) +26.2840 \quad {\rm (chip1),} \\
I = R - 0.0969 (B - R) +26.3331 \quad {\rm (chip2),} \\
I = R - 0.0969 (B - R) +26.5937 \quad {\rm (chip3),}
\end{eqnarray}

\begin{eqnarray}
\label{eq:B2Vcolor}
V = B - 0.160 (B - R) +26.350 \quad {\rm (chip1),} \\
V = B - 0.160 (B - R) +26.800 \quad {\rm (chip2),} \\
V = B - 0.160 (B - R) +26.192 \quad {\rm (chip3).}
\end{eqnarray}
Here the slopes were estimated by the using full data of all 3 chips, and
the offsets were estimated individually for each chip.

\begin{figure}
\includegraphics[angle=-90, scale=0.7]{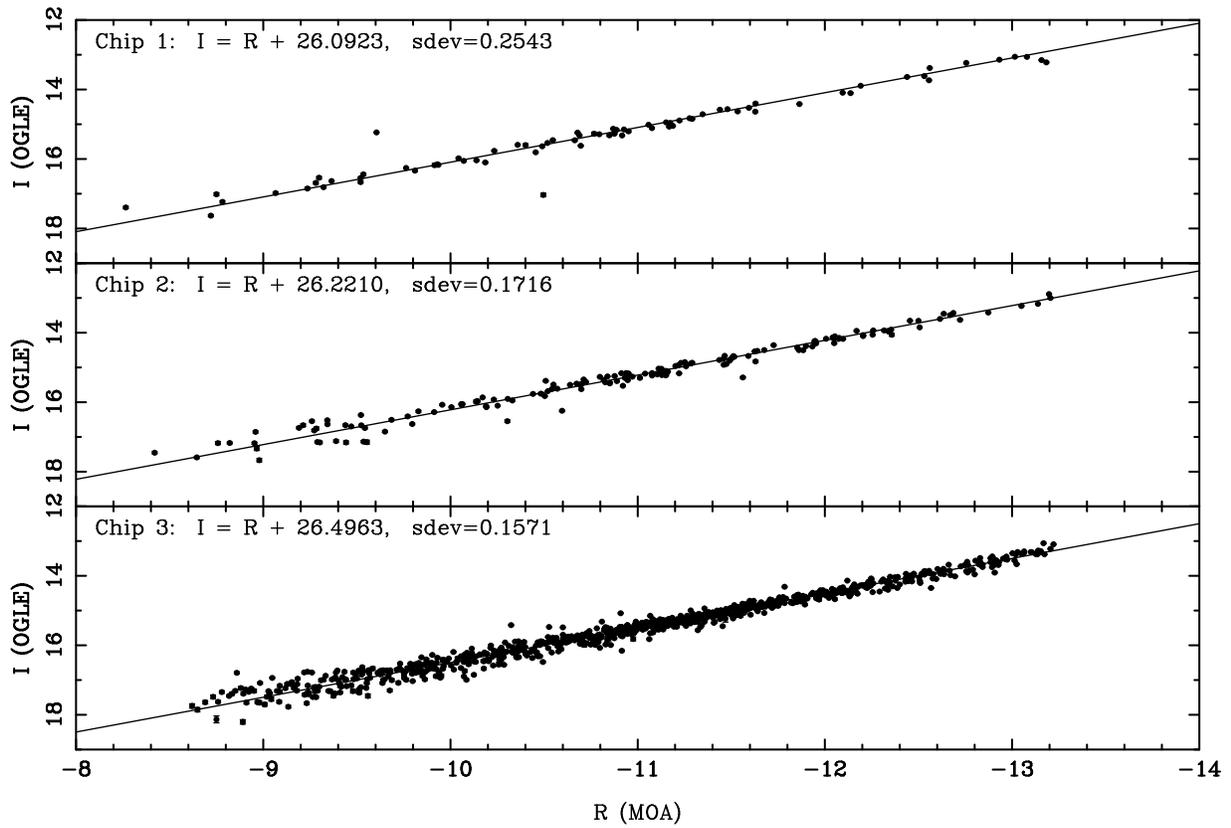}
\caption{Calibration of MOA Red measurements to the standard $I$ band
magnitude scale of the three CCDs of the camera MOA-cam2 without a
color term.}
  \label{fig:Rmed-I}
\end{figure}

We apply these transformations to our star catalogue.
As shown in Figure \ref{fig:Rmed-I_color}, the color dependence in transformation
to $I$ from $R$ is weak, and we find that the transformation without a
color term seems to work properly, as shown in Figure \ref{fig:Rmed-I}.
The transformation functions can be written as follows

\begin{eqnarray}
\label{eq:R2I}
I = R  + 26.0923 \quad {\rm (chip1),} \\
I = R  + 26.2210 \quad {\rm (chip2),} \\
I = R  + 26.4963 \quad {\rm (chip3).}
\end{eqnarray}

These equations are very useful because our data were usually
taken only in Red.

\section{Results}
\label{sec:results}

All data consist of $7,200$ images ($\sim 100$ GByte) which
correspond to $118 \sim 224$ frames for each field depending on
their priority. Of these, $6,600$ images ($\sim 92\%$) have been
successfully reduced. The reduction failures occur as a result
of poor data conditions, i.e., bad pointing, poor seeing and very
low transmission due to clouds.

In these subtracted images we find $2,000 \sim 10,000$ variable
objects in each field. And $200 \sim 1,000$ of these variable
objects pass the ``cut1'', for which the PSF photometry is 
performed to make the light curves. For these light curves, we
apply an additional selection, i.e., ``cut2'' which search for
a bump in the light curve and ``cut3'' in which microlensing
fitting is performed. As a result, we find $28$ microlensing
event candidates from these light curves.

\subsection{Event Selection}
\label{sec:event_selection}

We apply a combination of simple selection criteria to reject
the various spurious detections and variable stars. These criteria
are chosen empirically.

\subsubsection{cut1}
\label{sec:cut1}

We have applied a simple selection criterion termed ``point cut'' for
the initial variable objects detected in \S\,\ref{sec:detect}
by using the number of detections of each object in a time series
of frames.

There are 3 type of events in our data set. The first one is the
event in which the source star is not magnified in the reference
frame. The second one is that in which the source star is
magnified to its maximum amount relative to the reference image.
The third one is in-between, i.e., the source star is magnified
but not at the peak magnification relative to the reference image.
For type 1, objects are detected only when the source is magnified
as a positive excess. For type 2, objects are detected in almost
all frames except around the peak as a negative excess. And for
type 3, the same excess as for types 1 and 2, or the combination
of a cluster of positive excesses at the peak and a negative base
line are detected depending on the phase of magnification in the
reference image.

First we count the number of clusters of positive and negative detections
$N_{clus,p}$ and $N_{clus,n}$ which are the consecutive detections whose
separation are less than $4$ observation frames.

For type 1 which has the no negative detections $N_{det,n} =0$,
the positive detection should be larger than 2, i.e., $N_{det,p}\ge 2$, which
rejects many spurious detections and low S/N events.
$N_{clus,p}$ should be less than 4 to reject short period variables,
and $N_{det,p}$ should not be equal to $N_{clus,p}$ to reject objects
with noisy sparse detections.
We divide all frames into three regions of observation time, and calculate
the detection ratio $Ratio_{1,2,3}$ which is the number of detections
$N_{det,1,2,3}$ out of all observation frames $N_{frame,1,2,3}$.
We require that the $Ratio_1 \le 0.1$ or $Ratio_3 \le 0.1$ to reject some of
the long or middle period variables.

For type 2, which has the no positive detection $N_{det,p} =0$,
the negative detection should be larger than a quarter of all
observation frames, i.e.,  $N_{det,n}  \ge N_{frame}/4$ because
these negative detections represent the base line.
$N_{clus,n}$ should be less than 5 to reject short period variables
because the base line should be stable.

For type 3 which has both positive and negative detections,
$N_{clus,p}$ should be less than 4, and $N_{clus,n}$ should be less than 5
to reject short period variables, the same as for types 1 and 2.
Either of the criteria for type1 $N_{det,p} \ge 2$ and $N_{det,p} \ne N_{clus,p}$,
or for type2  $N_{det,n} \ge N_{frame}/4$ should be satisfied.
We list all criteria of cut1 in Table \ref{tbl:cut1}.

These cuts reduce the variable objects from thousands ($2,000 \sim
10,000$) to hundreds ($200 \sim 1,000$), depending on each field,
without rejecting the microlensing-like light curves. For these
objects the PSF photometry is performed to construct light
curves.

\subsubsection{cut2}
\label{sec:cut2}

The second cut ``cut2'' have been applied
to the light curves of objects which passed ``cut1''.
This cut search for a microlensing-like ``bump'' on the stable
baseline in the light curve.

First, we require that the number of photometric data points
$N_{total}$ should be $\geq 70$ points. Next we set the time
interval of $120$ days as a ``window'' in the light curve, where
the number of data points in the window $N_{in}$ and outside the
window  $N_{out}$ is required to be more than 3 and 9 points
respectively. In this window, we count the number of peaks
$N_{peak}$, which is defined to be that consecutive excess whose
significance is larger than $2.5 \sigma$ and at least the
significance of $2$ points of these should larger than $4 \sigma$.
Here the significance of each photometric point is calculated as
follows

\begin{equation}
  \label{eq:sigma_peak}
  \sigma_i = \frac{f_i -f_{med,out}}{\sqrt{\sigma_{f,i}^2 + \sigma_{out}}},
\end{equation}
where $f_i$ and $\sigma_{f,i}$ are the flux and error of the $i$th data point,
and $f_{med,out}$ and $\sigma_{out}$ are the median and variance of data
points outside the window.

$N_{peak}$ is required to satisfy $1 \le N_{peak} \le 3$,
and the maximum of the sum of the significance of the points in each peak
$\sum_{i,peak} \sigma_i$ should be larger than 20.
We define a high excess as one or
more consecutive data points whose significance is larger than
$2.5 \sigma$, i.e., all excesses including ``peak'' defined above and one
which is not significant to categorize as a ``peak''.
The number of the high excesses $N_{hi}$ in the window should be
less than 6, which rejects short period variables and
noisy light curves.
We require that the reduced chi-square of data outside the window $\chi^2_{out}$
should be less than 4. But if the ratio of chi-square inside and outside
the window $\chi^2_{in}/\chi^2_{out}$ is larger than 15, the cut with
$\chi^2_{out}$ is not required, so as to allow the high S/N long duration 
events to pass.  We list all these criteria of cut2 in Table \ref{tbl:cut2}.

In this ``cut2'', $24,543$ light curves have been reduced to $1,014$ in
all our GB fields.

\subsubsection{cut3}
\label{sec:cut3}

For the light curves which passed cut2, the microlensing model
fitting with equation (\ref{eq:amp-u}) have been applied in cut3. In
DoPHOT analysis the fittings are performed with the background
blending flux $F_b$, i.e.,  $F(t)= F_0A(t) + F_b$, where $F_0$
indicates a baseline source flux. However, in DIA, we observe only
the variation of the flux from that in the reference image as
follows

\begin{equation}
  \label{eq:deltaf}
  \Delta F(t)= F_0A(t) - F_0A(t_{ref}),
\end{equation}
where $t$ and $t_{ref}$ are the time when the current and the reference image
are taken respectively. $F_b$ is canceled out in this formula.
The equation (\ref{eq:deltaf}) with four parameters ($F_0, t_0, u_{\rm min}, 
t_{\rm E}$) is fitted to differential flux $\Delta F$ light curves in this cut3.

The reduced chi-squares $\chi^2$ in a microlensing fit are
required to be less than 3.5 to reject most of LPVs and noisy
light curves, although very high S/N events are sometimes not well
fitted by the standard microlensing model because of exotic
effects such as parallax. So, for the events whose peak flux is
larger than $450,000$ ADU, we require  $\chi^2 <100$ instead of
$\chi^2 <3.5$. The event ngb1-2-2717 falls into this category
because of the parallax effect. After these cuts, 75 light curves
still remain, most of which have a clear single peak and stable
baseline. These are microlensing candidates, dwarf novae (DN) and
low S/N faint LPVs.

The main background in this event selection are DN which 
can be well fitted by microlensing in the case of poor sampling.
In our light curves which passed the cut1, 20 light
curves are clearly identified as DN.
These have single or sometimes multiple asymmetric flares in the
light curve, which usually rise quickly and fade slowly.

To check whether an object is DN or not, we cross-referenced to
the existing light curves in the MACHO group's database
(http://wwwmacho.mcmaster.ca/) whose fields overlap some of ours,
and found other flares in the light curves of 5 objects, i.e.,
these are clearly DN. The DN are much bluer during the
outburst. However we usually observe only with the red filter to
increase the sampling rate, and only rarely observe with the blue
(only $\sim 30$ frames for each field). For 23 out of these 75
candidates, we could measure the color during the flare. Five of
these 23 candidates are categorized as DN because they are very
blue ($V-I \le 1.0$). In 15 light curves, including 5 of the above
10 DN, asymmetric flares are clearly seen. In total 20 objects
are categorized as DN.

Two of these DN are rejected at cut2. And one DN is removed
at $\chi^2$ by cut3. To reject other DN, we are unable to use
a color cut because we have no color information in most of the
light curves. On the other hand we have frequent sampling data
points for these objects, which makes it easier to identify a DN 
from the shape.  And we find the following cuts would reject
almost all DN in addition to LPVs.

The minimum impact parameter $u_{\rm min}$ should be less than 1. 15
DN are fitted with a large $u_{\rm min}$. Some of LPVs are also
cut here. Furthermore we impose the condition that the timescale
should be $0.3<t_{\rm E} < 200$ days. Two further DN are rejected
here, which are fitted to very large event timescales ($t_{\rm E}
>1000$ days). In consequence, all DN which have been clearly
identified are rejected by these cuts. Though 5 of these
have been identified as DN not by a clear shape of the flare but by
their color or by the existence of other flares in past data,
which can be done randomly,
these are rejected by our cuts. These results give us the
confidence that all DN are rejected by these cuts. Even if a
few DN are not rejected, it should not be a significant
fraction in our results. At the same time 9 microlensing-like
light curves, which could not be clearly identified as either
DN or microlensing events, are rejected with this cut.  In
microlensing events measured only around the peak or with low S/N,
$t_{\rm E}$ and $u_{\rm min}$ would be degenerate (\citealt{han00}).
So some real microlensing events might be removed at this cut.
These effects are also seen for the artificial events with faint
source stars in our simulation (see \S\,\ref{sec:efficiency}).
If both the colors are taken simultaneously or the catalogue of
DN is used, we can easily distinguish real microlensing events
from DN and issue the alerts. Such events are usually high
magnification events because their source star is very faint. If
the baseline flux is measured by follow up observations with
larger high resolution telescope such as HST or VLT, the timescale
is well constrained. Other LPVs are also cut here.

31 objects have passed these criteria. In these light curves 3 low S/N
LPVs still remain. We have rejected these LPVs directly as we are 
doing in real-time analysis (\citealt{bon01}) instead of imposing
more complicated criteria. In real-time analysis we have made a list of
variables and those objects which cross-referenced to these
variables are rejected. This treatment does not affect the
optical depth estimation because that is position-dependent and
these positions are rejected in following analysis. We list all
the criteria of cut3 in Table \ref{tbl:cut3}.

\begin{figure}
\plotone{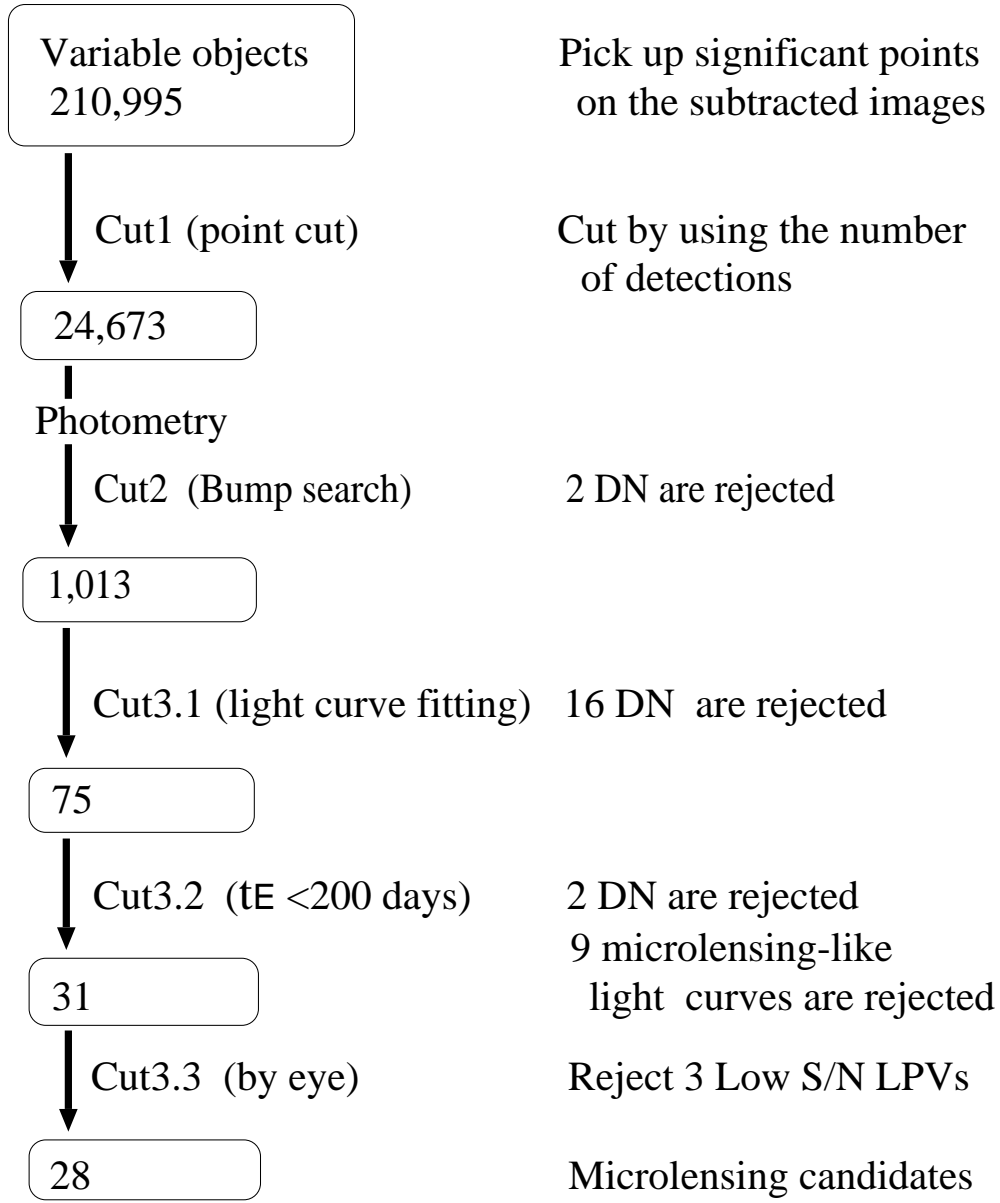}
\caption{A flowchart of the procedure for selection of microlensing candidates.}  \label{fig:flowchart}
\end{figure}

As a result, we find $28$ microlensing candidates in our GB
database during 2000.  We confirmed that these 28  candidates
didn't have any significant variability during the following two 
seasons (in 2001 and 2002).  Three out of 20 candidates reported in
real-time analysis (\citealt{bon01}) have failed in this off-line
analysis because the $t_{\rm E}$ are not well constrained. And 11 new
candidates are found because we have changed the threshold to detect
variable objects on the subtracted images in \S\,\ref{sec:detect}.
We summarize the event selection processes in Figure
\ref{fig:flowchart}.

\begin{figure}
\epsscale{0.9}
\plotone{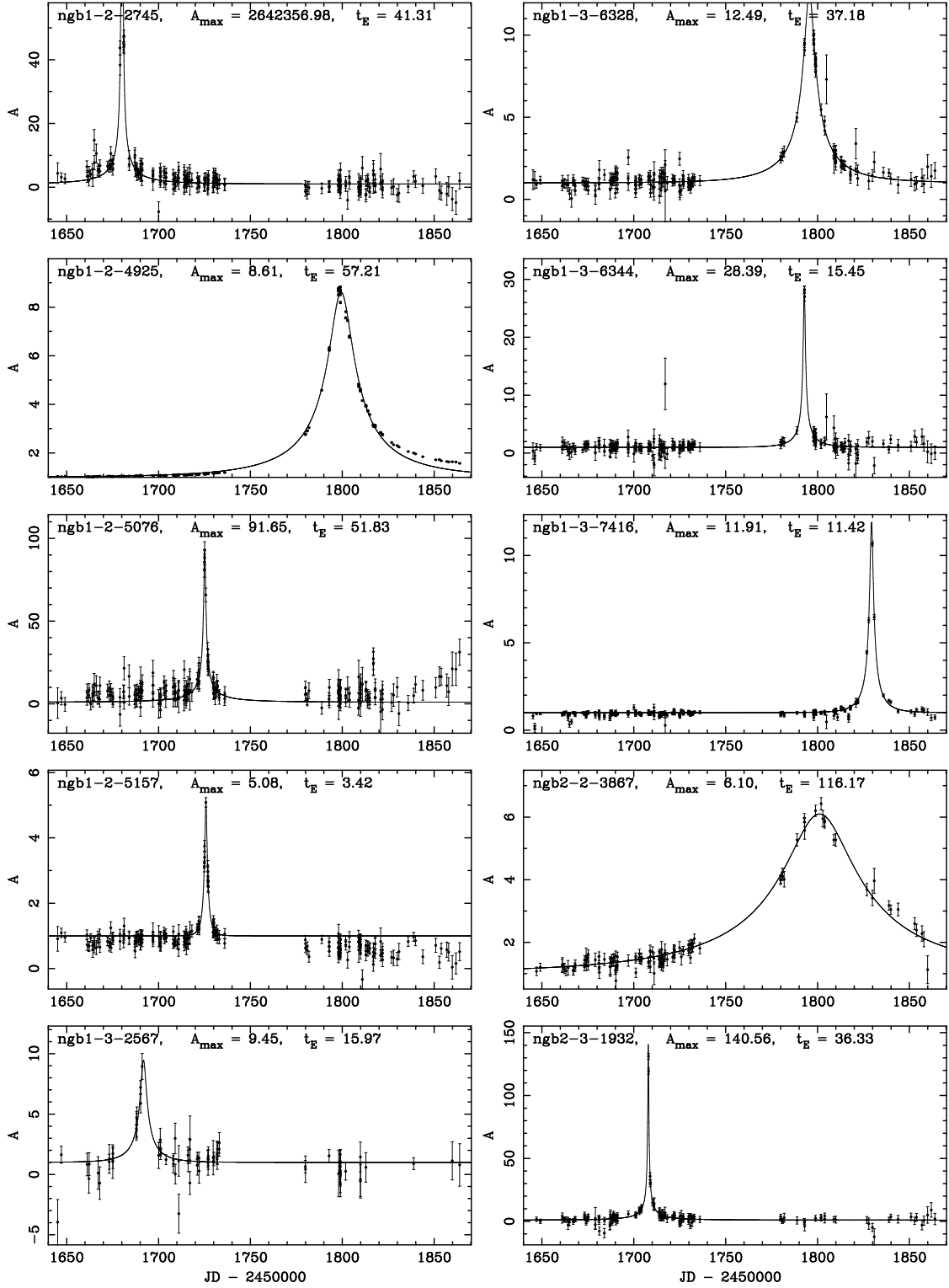}
\caption{Light curves of microlensing event candidates.}
\label{fig:lc1-10}
\end{figure}

\begin{figure}
\plotone{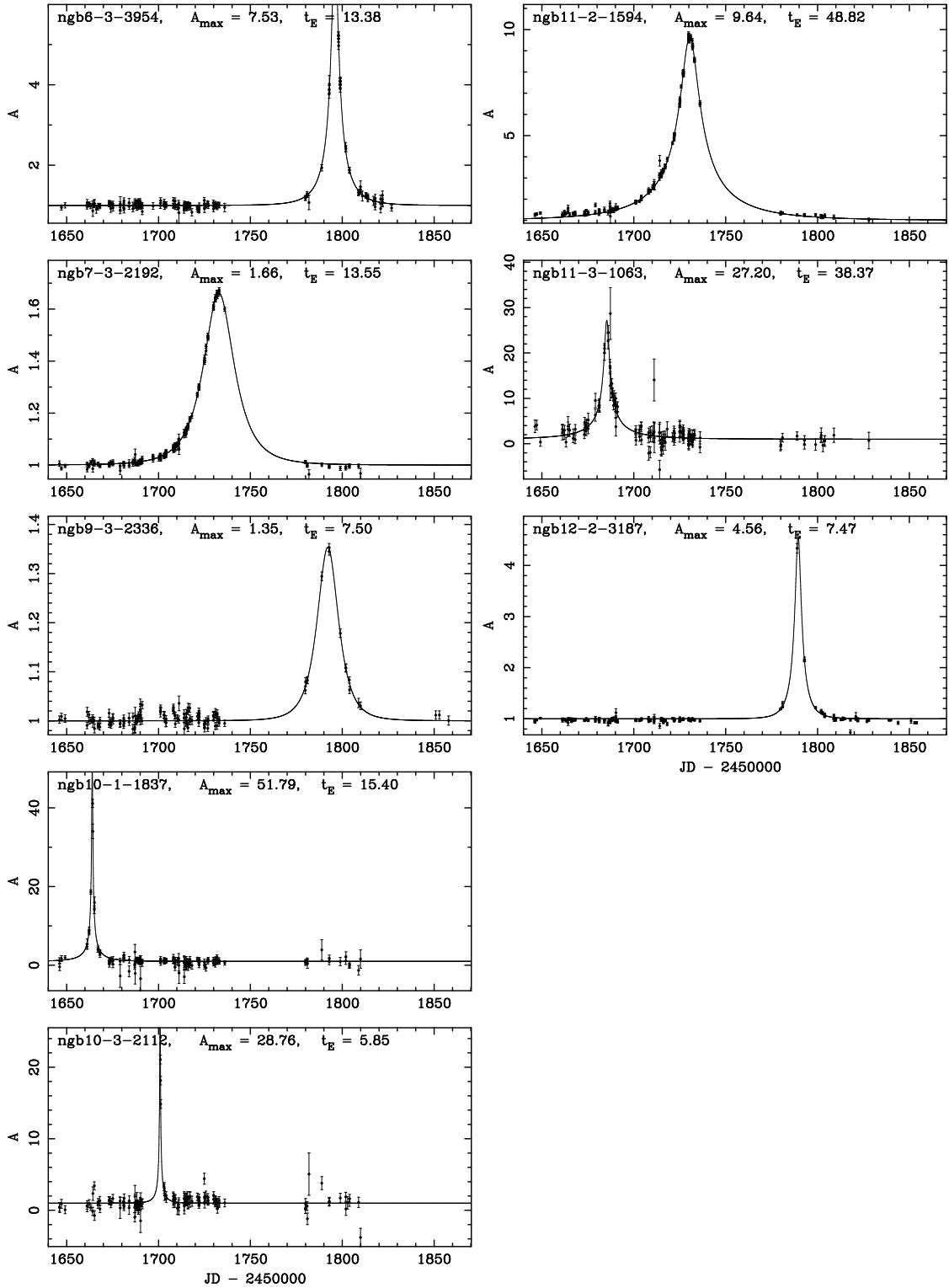}
\caption{Light curves of microlensing event candidates.}
\label{fig:lc21-28}
\end{figure}

We show light curves of these 28 candidates in Figure \ref{fig:lc1-10}
$\sim$ \ref{fig:lc21-28}, where the $\Delta F$ data points have been
converted to amplifications using the fitted parameters.
The gap around JD $= 2451750$ in the light curves is due to
the mal-functioning of our camera system for $\sim 40$ days.

We list the positions of all candidates with ID in this analysis
in Table \ref{tbl:candidates}.  The ID in real-time analysis and
alert ID  reported in \cite{bon01} are also tabulated.  We list
the best fit microlensing parameters and $1 \sigma$ lower and
upper limits in Table \ref{tbl:parameters}. The $I$-band baseline
magnitude of the source star $I_0$ is de-reddened to match the HST
field by using the $I$-band extinction $A_I$ map of each field.
These fitted parameters are not biased by the blending effects due
to nearby stars which appear in DoPHOT-type analyses
(\citealt{uda94}; \citealt{alc97a,alc97b})and would make $t_{\rm
E}$ shorter. In DIA all other blending components could be
subtracted.

The light curve of the event ngb1-2-4925 exhibits an asymmetric
profile due to the effect of parallax. The best fit parameters in
fitting with parallax microlensing models for this event are
presented in \cite{bon01}. We will use these values for the
following optical depth estimation.

\section{Extinction Map}
\label{sec:extinction}

\begin{figure}
\epsscale{0.5}
\includegraphics[angle=-90, scale=0.6]{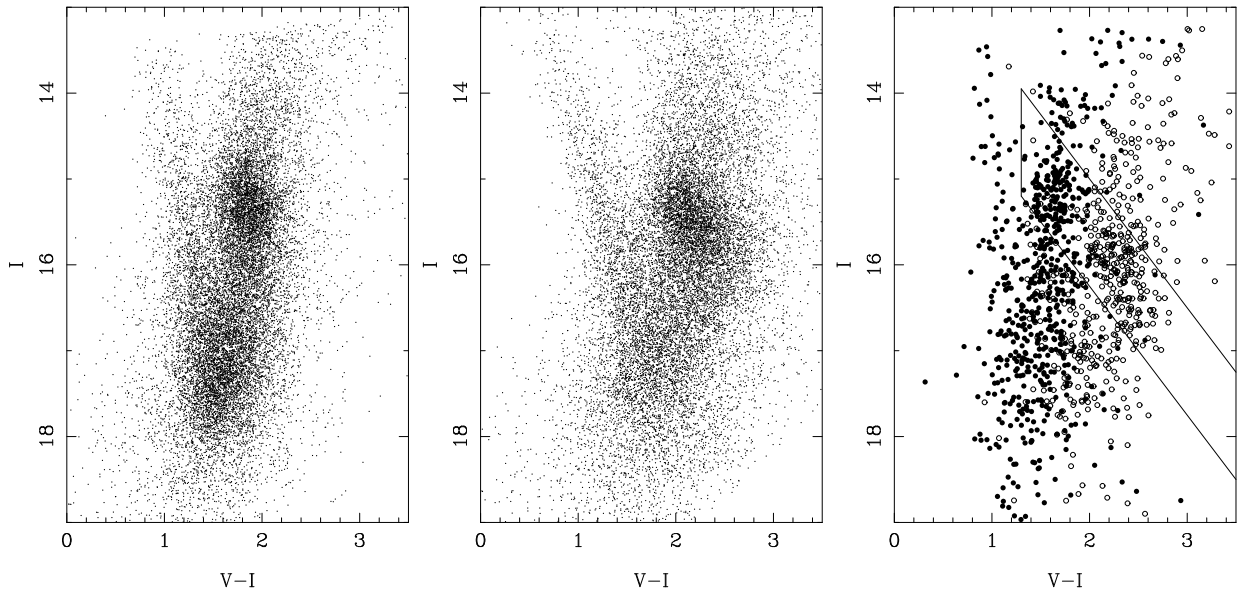}
\caption{CMD of the Baade's window field (left panel),
a higher extinction field (middle panel) and two subregions
in the Baade's window field (right panel). The extinction $A_I$ in each subregion
(in right panel) are 0.761 (filled circle), 1.645 (open circle).
The box in the right panel encloses the red clump giant region defined by
$I - 1.5 (V-I) = 1.2, 13.25$.
\label{fig:cmd}}
\end{figure}

As is well known, the extinction due to dust is very significant
towards the GB. This effect can be seen in our Color
Magnitude Diagram (CMD) of the field near Baade's Window (left
panel) and another high extinction field (middle panel) in Figure
\ref{fig:cmd}. The CMD is more scattered in the high extinction
field than in the Baade's Window field due to the extinction and
the reddening. Information on the extinction in each region is
needed to estimate the number of stars in this region by using the
luminosity function.

We apply \cite{pac99}'s method which use red clump-dominated
parts of the CMDs for determining the offsets caused by
differential extinction. They have made a reddening map for their
fields because determining the reddening $E_{V-I}$ (horizontal
shift in CMD) is easier than $A_I$ (vertical shift in CMD).  Then
the extinction map are calculated by the following formula
(\citealt{woz96}; \citealt{ng96}; \citealt{sta96}):

\begin{eqnarray}
  \label{eq:ai}
  A_I =  1.5 \times E_{V-I}.
\end{eqnarray}

We have made CMDs of $I$ vs $(V-I)$ for each of $8 \times 16$
sub-divided regions ($3.45' \times 3.45'$) by using our star
catalogue. A sample of these CMDs of different subregions in the
image near Baade's window is shown in the right panel of Figure
\ref{fig:cmd}.  In this figure, differential reddening is clearly
seen. We select the red-clump-giant region in the CMD following
\cite{pac99} as indicated in Figure \ref{fig:cmd}. The parallel
lines whose slopes are the reddening vectors (\citealt{sta96}) are
given by

\begin{equation}
  I - 1.5 (V-I) = 1.2, 13.25.
\end{equation}

\begin{figure}
\epsscale{0.8}
\includegraphics[angle=-90, scale=0.7]{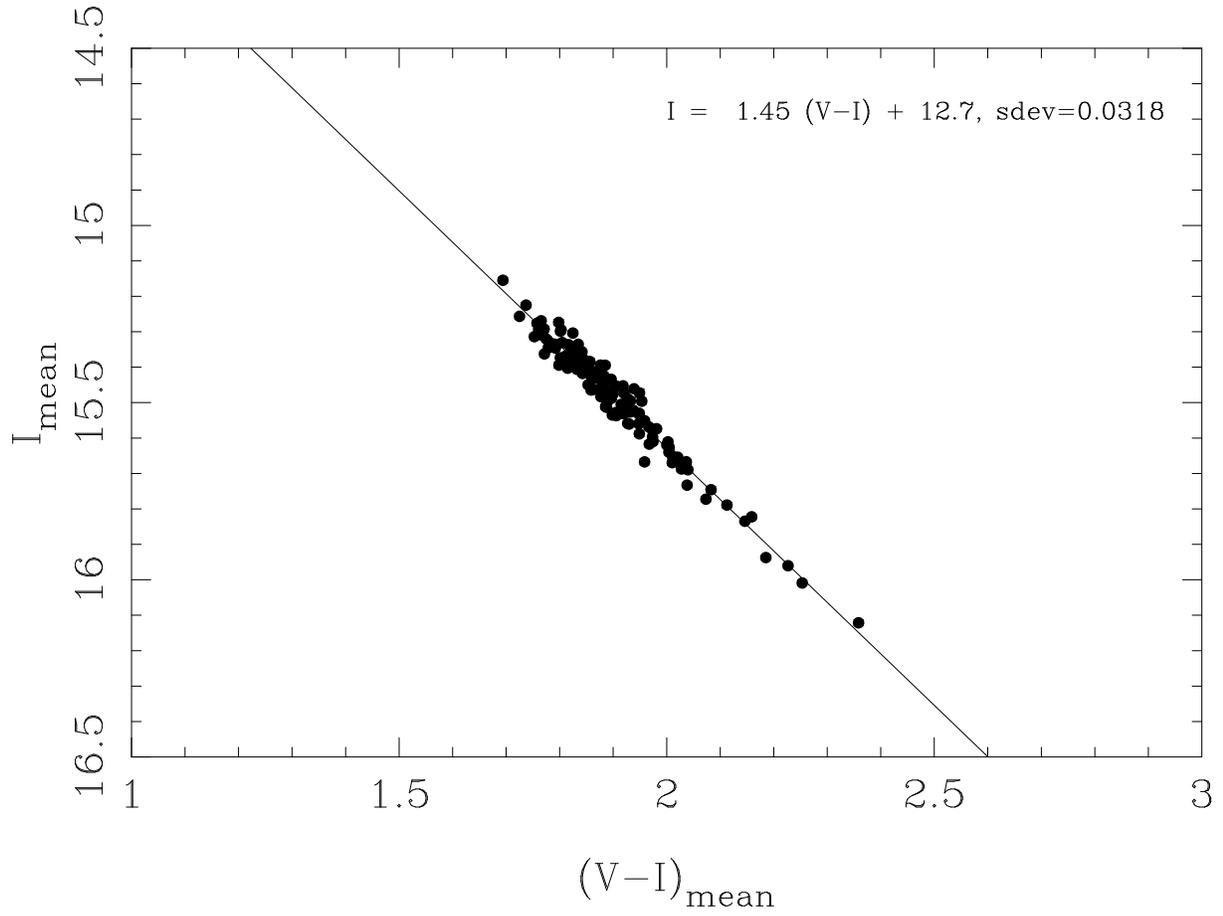}
\caption{ Correlation between the mean of  $(V-I)$ and $I$ for red
clump giants in each subregion.
The fitted line is $I = 1.45 (V-I) + 12.7$.
\label{fig:reddning_curve}}
\end{figure}

\begin{figure}
\includegraphics[angle=-90, scale=0.65]{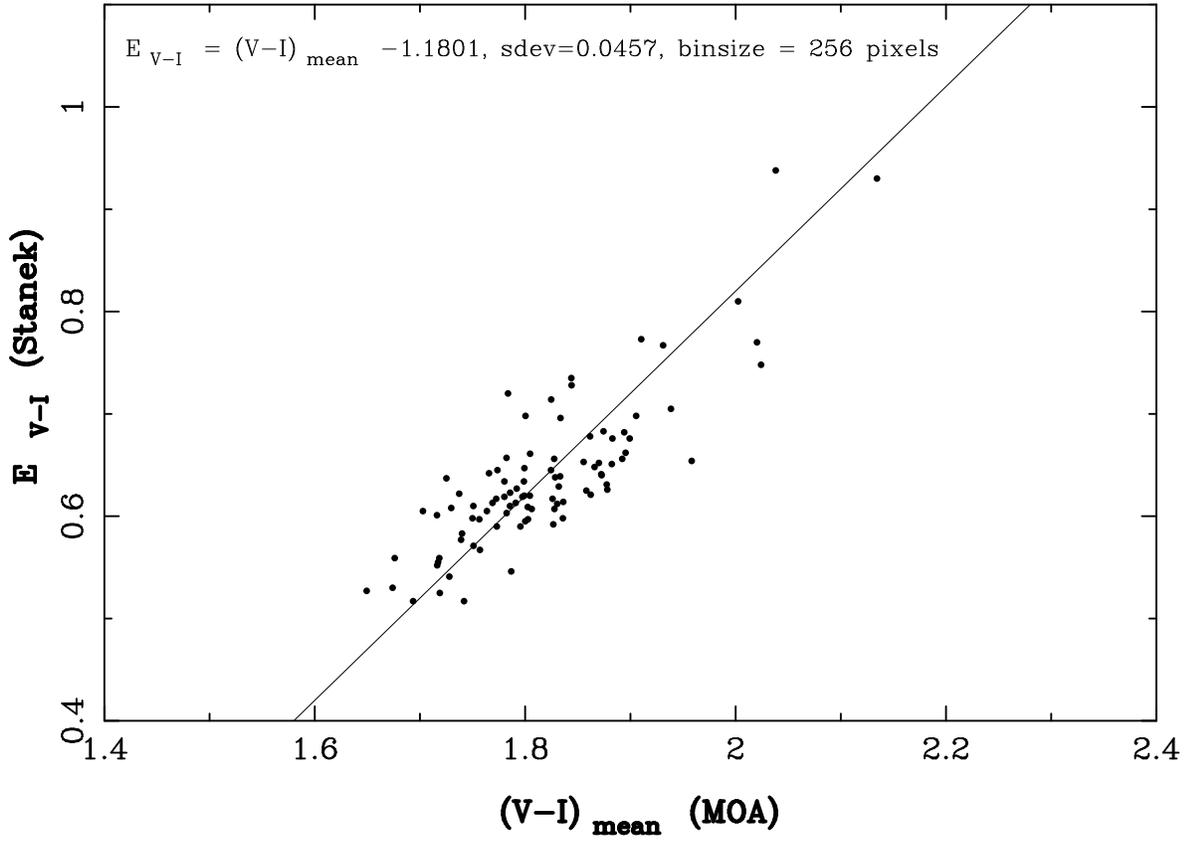}
\caption{ Correlation between $(V-I)_{mean}$ from MOA and $E_{V-I}$
from \cite{sta96} in the overlap regions.
The fitted line is $E_{V-I} = (V-I)_{mean} -1.18 \pm 0.05$.}
\label{fig:color_EVI}
\end{figure}

We estimate the mean values of $(V-I)$ and $I$ for red
clump giants in each subregion, which correspond to the correlation between
differential reddening $E(V-I)$ and differential extinction $A_I$.
These values in a field are plotted in Figure \ref{fig:reddning_curve},
where the best fitted line is given by

\begin{equation}
   I = (1.45 \pm 0.12) (V-I) + 12.7.
\end{equation}
This slope is consistent with \cite{sta96}. The mean colors
$(V-I)_{mean}$ is more reliable than $I$. So we estimate
$E_{V-I}$ from $(V-I)$ as follows. We compare  $(V-I)_{mean}$
from our data  and the reddening $E_{V-I}$ map (centered on
(18:03:20.9, $-30$:02:06)) calculated by \cite{sta96} in the
overlap regions (ngb2-1, ngb2-2 and ngb3-3), where the zero point
is based on the determination by \cite{goupop98} and \cite{alc98}.
A correlation between $(V-I)_{mean}$ and $E_{V-I}$ is shown in
Figure \ref{fig:color_EVI} and equation (\ref{eq:VI_EVI}).

\begin{equation}
\label{eq:VI_EVI}
   E_{V-I} = (V-I)_{mean} -1.18 \pm 0.05.
\end{equation}

\begin{figure}
\includegraphics[angle=-90, scale=0.6]{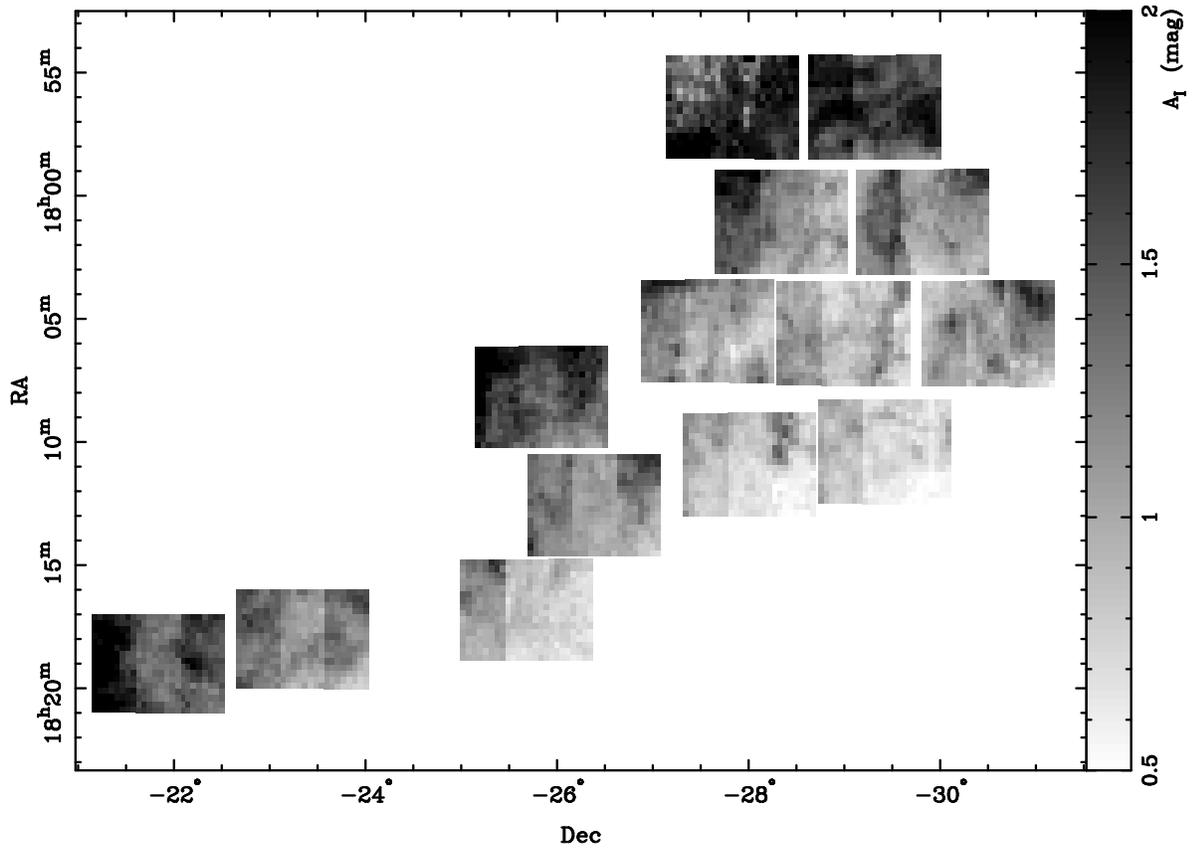}
\caption{ $I$-band extinction $A_I$ map of 14 Galactic Bulge
fields. The darker regions indicate higher extinction,lighter
regions mean lower extinction. We can see the galactic disk, in
which the extinction is high, in the diagonal line from the
top-right to bottom-left. }
\label{fig:AI_map}
\end{figure}

Using equations (\ref{eq:ai}) and  (\ref{eq:VI_EVI}), we have made an 
$I$-band extinction $A_I$ map from the $(V-I)_{mean}$ map in our data
as shown in Figure \ref{fig:AI_map}. The error in $A_I$ is about $0.07$ 
mag. This map is used to make a luminosity function for each region.

\section{Optical Depth}
\label{sec:OPTICALDEPTH}

To determine the microlensing optical depth from detected events
we estimate our detection efficiency by Monte Carlo simulations.
Here we add artificial microlensed star images to the real
subtracted images instead of a full Monte Carlo approach. Then we
count the number of events which passed the same event selection
applied in the real event selection. Then we estimate the optical
depth towards the Galactic bulge.

\subsection{Luminosity Functions}
\label{sec:lf}

To add the artificial microlensing events, we need
the Luminosity Functions (LF) of source stars in GB fields.
We use the deepest observations with the {\it Hubble Space Telescope} (HST)
from \cite{hol98}, which  measured stars in Baade's window down to
$I$ (F814W) $\sim 24$.

A completeness correction has been applied for the HST LF for
fainter stars. The bright end of the HST LF is poorly defined
because of the small field of view ($\sim 5$ arcmin$^2$).
Meanwhile the MOA data have a lot of bright stars because of the
wide field of view ($\sim 1.2$ deg$^2$), but are poor for fainter
stars. So we combine the HST LF and MOA LF at $I = 15 \sim 16$ as
shown in Figure \ref{fig:lf}. Here the MOA LF is made from stars
in field ngb2-2 ($0.4$ deg$^2$) which includes the HST field. The
offsets in the photometry of stars in each subregion due to
differential extinction are corrected to match that of the HST
field ($A_I = 0.742$ mag). We use the MOA LF for $I<16$ and the
HST LF for $I \ge 16$ in this analysis. We use this composite LF
for all our GB fields assuming that the morphology of the LF
varies little between the small HST field and the large MOA
fields.

\subsection{Simulation}
\label{sec:simulation}

In each of our 42 ($14$ fields $\times$ $3$ chips) subfields, the
density of stars, sampling rate, and observational conditions are
different. So we estimate the detection efficiencies individually
for each subfield by a Monte Carlo simulation. We generate
429,000 artificial microlensing events in each subfield for this
purpose. 

A given event has 4 parameters: ($u_{\rm min},t_{\rm E}, t_0, I_0$).
We generated artificial events with the timescale
$t_{\rm E} = (0.3, 0.5, 1,3, 5, 10, 20, 40, 60, 80, 100, 150, 200)$ days.
For each $t_{\rm E}$, we generated events with source extinction-corrected 
$I$-band magnitude between 12.0 $\le I_0 \le$ 22.9 mag at intervals of 0.1 mag 
uniformly (the corresponding flux values $F_0$ were estimated by taking 
the extinction at each position into account). For each value of 
$t_{\rm E}$ and $I_0$,  300 events were generated (i.e. 429,000 events 
for each subfield).

Those parameters which follow the well known distribution such
as the position $(x,y)$, the peak time $t_0$ and the impact
parameter $u_{\rm min}$ are selected at random uniformly in the CCD chip 
($36 \le x \le 2042$ pixels and $5 \le x \le 4090$ pixels),
in the observation period ($2451645.1\le JD\le2451863.9$) and 
in $0<u_{\rm min}<u_{th}$, respectively.
Here, $u_{\rm th} = 10^{-0.4(I_0-I_{\rm th})} (I_0 \ge I_{\rm th}$) and
$u_{\rm th}=1$ ($I_0 \le I_{\rm th}$).
We set $I_{\rm th} = 16$ mag (for ngb1,4,5 and 10), 16.5 mag 
(for ngb2,3,6,8 and 11) and 17 mag (for ngb7,9 and 12) depending on 
the mean extinction in their fields.
We ensured that the events with $u_{\rm min}>u_{\rm th}$ are not detectable
in our experiments from former simulations.
This $u_{\rm th}$ is introduced to enlarge the statistics for the faint source
events.  The slope of $1/u_{\rm th}$
is roughly consistent with the slope of our combined GB LF, $\psi(I)$,
which means the number of generated events is roughly following the LF for
faint source events.

 Many other components which are
expected to affect the event detection, e.g., bad pixels, seeing,
extinction, sample rate, star density, focusing, tracking, sky
background, clouds, PSF variation in the focal plane, differential
refraction, variable stars, saturated stars, cosmic rays,
satellite, asteroids, systematic residuals and any other unknown
systematic noise, are automatically simulated because we use
real subtracted images in this simulation.


To generate artificial microlensing events, we add the
artificial differential star images on a time series of real
subtracted images. The differential stars are made from the local
PSF which was formed by the same procedure outlined in 
\S\,\ref{sec:photometry}, and scaled by the differential flux
expressed in equation (\ref{eq:deltaf}).

In generating the artificial image, the photon noise is taken
into account following Poisson statistics with an additional flux
in the current image. The ADU flux of the artificial star in the
current image $F_0A(t)$ in equation (\ref{eq:deltaf}) is 
calculated in the scale of the reference image. So, to estimate
the photon noise from the additional flux, this additional flux
should be transformed to the number of electrons in the current
image given by $F_0 a_0 /G$, where $G$ and $a_0$ are the gain in
ADU$/e^-$ and a scale factor in the kernel respectively. Then the
electron signals are simulated randomly following a Poisson
distribution with the mean value of $F_0 a_0 /G$. The resultant
simulated random electron signals are rescaled back to the
reference image in ADU by multiplying by $G/a_0$.


The photon noise from the sky and the blending stars in the current 
image and reference image are already included in the real subtracted
images. Furthermore the systematic noise from the poor subtraction
due to the low S/N is also included. These noise sources heavily 
depend on the sky condition, seeing, tracking of the telescope and 
star density  of each frame. So it is complicated to simulate with 
full artificial images.

\begin{figure}
\epsscale{0.5}
\plotone{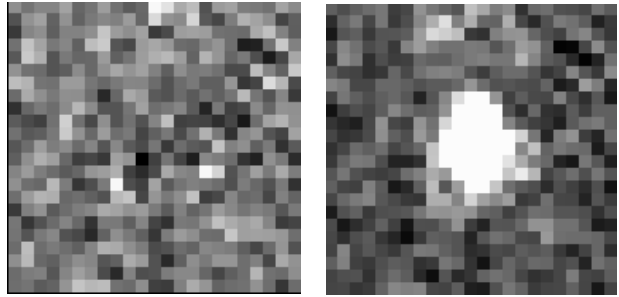}
\caption{Sample of a cut raw subtracted subimage ($23 \times 23$ pixels)
without (left) and with (right) an artificial event.}
  \label{fig:artimage}
\end{figure}

We cut out a small sub-image of $23 \times 23$ pixels from the
subtracted image at a randomly selected position,  and put the
corresponding differential flux $\Delta F(t)$ onto each sub-image.
A sample of the cut raw subtracted image (left panel) and the
generated artificial differential image (right panel) are shown in
Figure \ref{fig:artimage}.

\begin{figure}
\epsscale{0.8}
\plotone{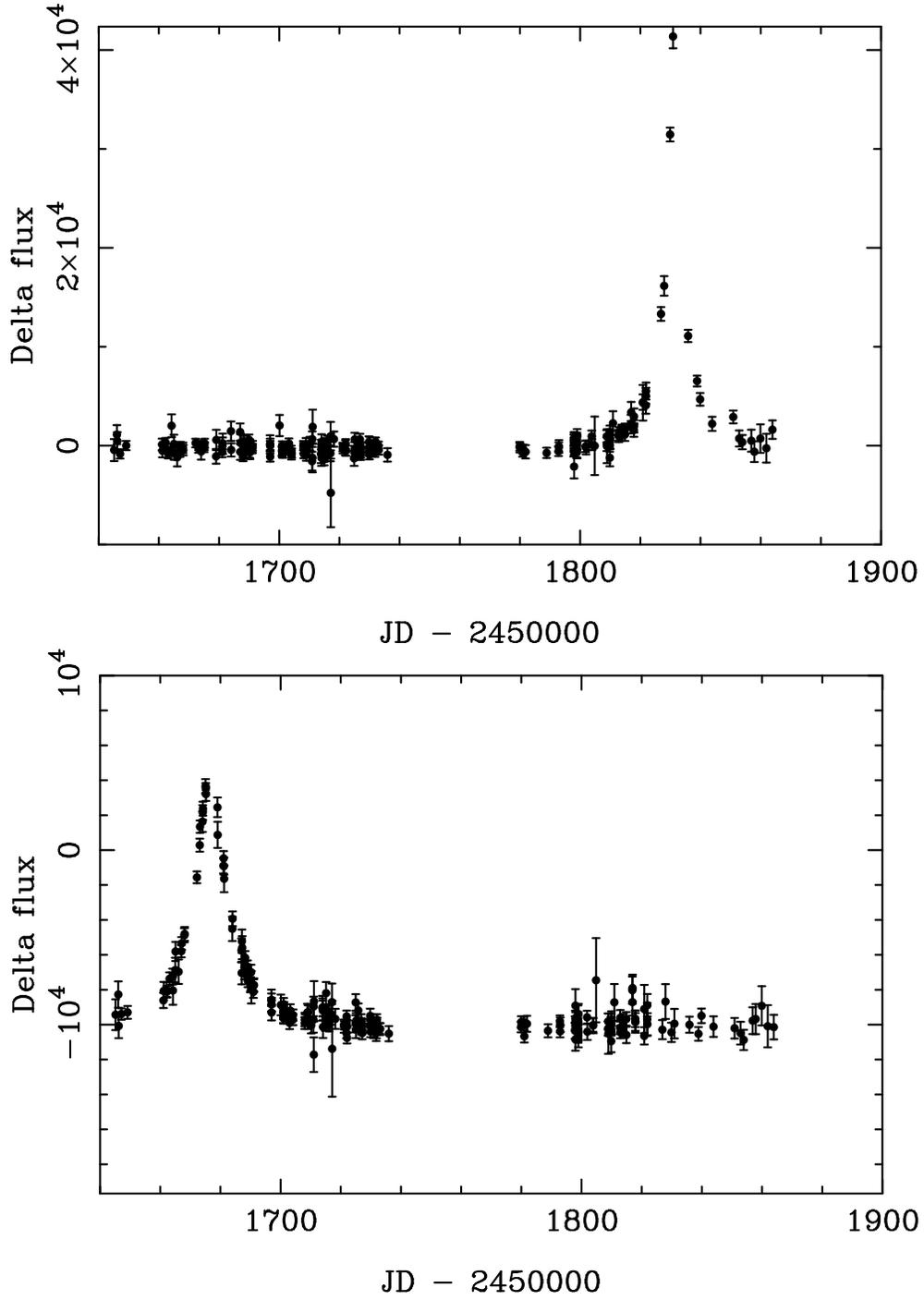}
\caption{Sample light curves of artificial microlensing events
with $t_{\rm E} = 20$ days and $I_0 \sim 17$.}
\label{fig:art_lc}
\end{figure}

For a series of generated artificial images, we apply the same
variable object detection process as in \S\,\ref{sec:detect}, and
the same PSF photometry as in \S\,\ref{sec:photometry}. Saturated
pixels ($> 30,000$ ADU) are rejected as in the real analysis. The
resultant light curves are stored in the database of artificial
events. Sample artificial light curves are shown in Figure
\ref{fig:art_lc}.

We count the number of artificial events $n(t_{\rm E}, I_0)$ which passed all
criteria for the real event selection for each $t_{\rm E}$ and $I_0$. Then we 
calculate our detection efficiencies.

\subsection{Detection Efficiencies}
\label{sec:efficiency}

We show our detection efficiencies as a function of
the event timescale, the source magnitude and the minimum impact
parameter.

\subsubsection{Detection Efficiency as a function of Event Timescale}
\label{sec:eff}

The detection efficiency $\varepsilon_0(t_{\rm E}, I_0)$ for each sample
with $t_{\rm E}$ and $I_0$  is given by
$\varepsilon_0(t_{\rm E}, I_0) =  n(t_{\rm E}, I_0)/300$,
where $n(t_{\rm E}, I_0)$ is the number of artificial events which passed
our event selections. Then efficiency $\varepsilon(t_{\rm E})$ for $t_{\rm E}$ 
is integrated by weighting by the LF $\psi(I)$ as follows,

\begin{equation}
  \label{eq:eff}
    \varepsilon(t_{\rm E})=\frac{1}{N_{23}}\int_{I=12}^{I=23} 
        \varepsilon_0(t_{\rm E}, I) \psi(I) u_{\rm th} dI,
\end{equation}

where 

\begin{equation}
  \label{eq:eff2}
N_{23} = \int_{I=12}^{I=23}  \psi(I) dI.
\end{equation}

\begin{figure}
\includegraphics[angle=-90, scale=0.7]{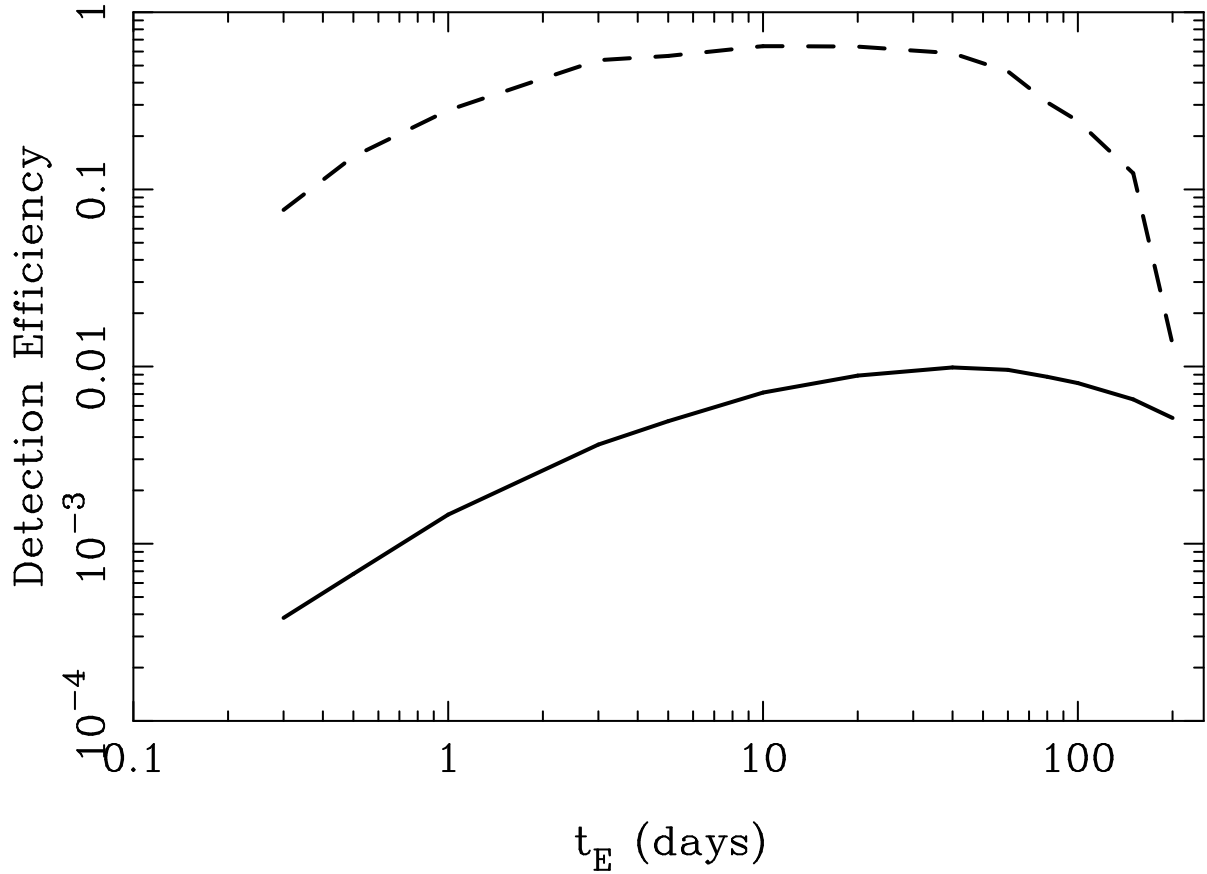}
\caption{MOA detection efficiency as a function of $t_{\rm E}$ for
the ngb1-2 subfield for all source stars ($I < 23$, solid line) and
for bright sources ($I = 14$, dashed line).  The efficiencies slightly
differ for each subfield because of differences in the sampling rate,
the star density and the extinction.}
\label{fig:eff}
\end{figure}

We show one of the resultant detection efficiencies as a function
of $t_{\rm E}$ for bright stars ($I \sim 14$, dashed line) and
for all possible sources ($I <23$, solid line) in Figure \ref{fig:eff}.
Most events with bright source stars whose timescale is longer than $3$
days can usually be detected.

On the other hand, the total event detection efficiencies (solid
line, $I<23$) are very low. This is because most source stars are
fainter than our observational limiting magnitude of $I\sim 17$.
These faint source stars can be detected by microlensing only when
they are highly magnified. The possibility that such a high
magnification (i.e. small impact parameter $u_{\rm min}$) event occurs
is small.  Furthermore, such faint source events cannot be
detected any more for shorter timescale events.

\subsubsection{Detection Efficiency as a function of source magnitude}
\label{sec:eff_I}

\begin{figure}
\includegraphics[angle=-90, scale=0.7]{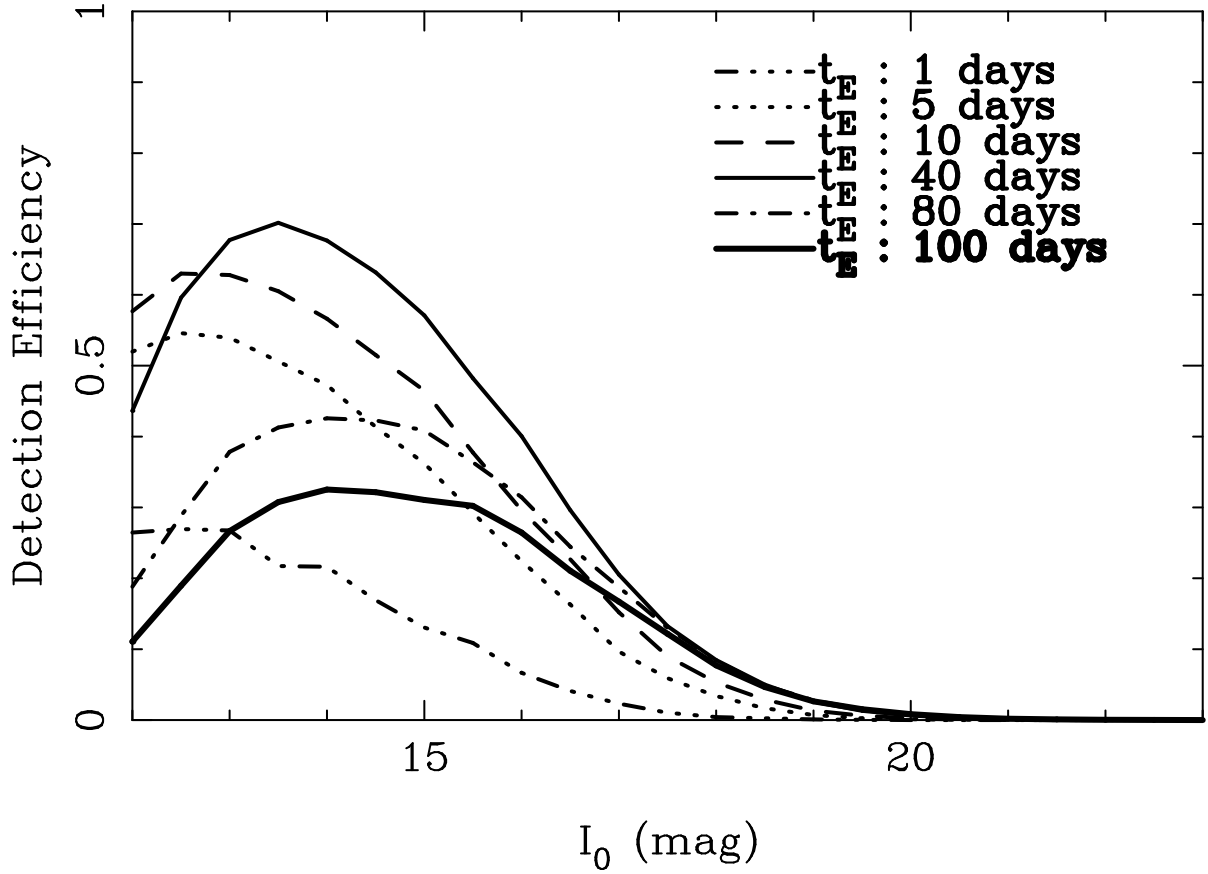}
\caption{MOA detection efficiencies averaged over all subfields as a function 
of source $I$-band magnitude $I_0$ for the various event timescales
as indicated in Figure.}
\label{fig:eff_I}
\end{figure}

We present the detection efficiencies averaged over all subfields 
as a function of the source 
$I$-band magnitude $I_0$ for various event timescales in
Figure \ref{fig:eff_I}, where  $I$-band magnitudes are extinction
corrected to match those of the HST field ($A_I= 0.742$).
The efficiencies are very high for brighter source events, as
expected. One can see the fall off at the bright end of this
efficiency curve due to the saturation of the source star images.
The fact that this effect appears in our simulation is one of the
pieces of evidence that this simulation is realistic. On the other
hand efficiencies fall off for fainter stars as expected. The
maximum of our detection efficiency is not high, even for
bright source events, because our data have a gap for $\sim 40$
days due to the mal-functioning of our camera system.

Comparing each of the timescales,  for the long timescale events,
the efficiency falls off more significantly  than that of the
shorter timescale events at the bright end of these curves. This
is because the number of saturated points is large  and we
required a stable flat baseline to identify the event.  In the
long timescale events ($t_{\rm E}>50$ days), the stable flat
baseline in the light curve is very short. These effects are
significant for brighter source and longer timescale events and
vice versa. Meanwhile, at the faint end, the efficiencies are
inverted compared with those at the bright end. For long timescale
events the duration with a significant excess in the light curve
is longer than that of  shorter timescale events which cannot be
detected any more. We note there is a small but still non-negligible 
efficiency down to $I\sim 22$.  This is the main improvement in the 
analysis by changing from DoPHOT to DIA.

\begin{figure}
\includegraphics[angle=-90, scale=0.65]{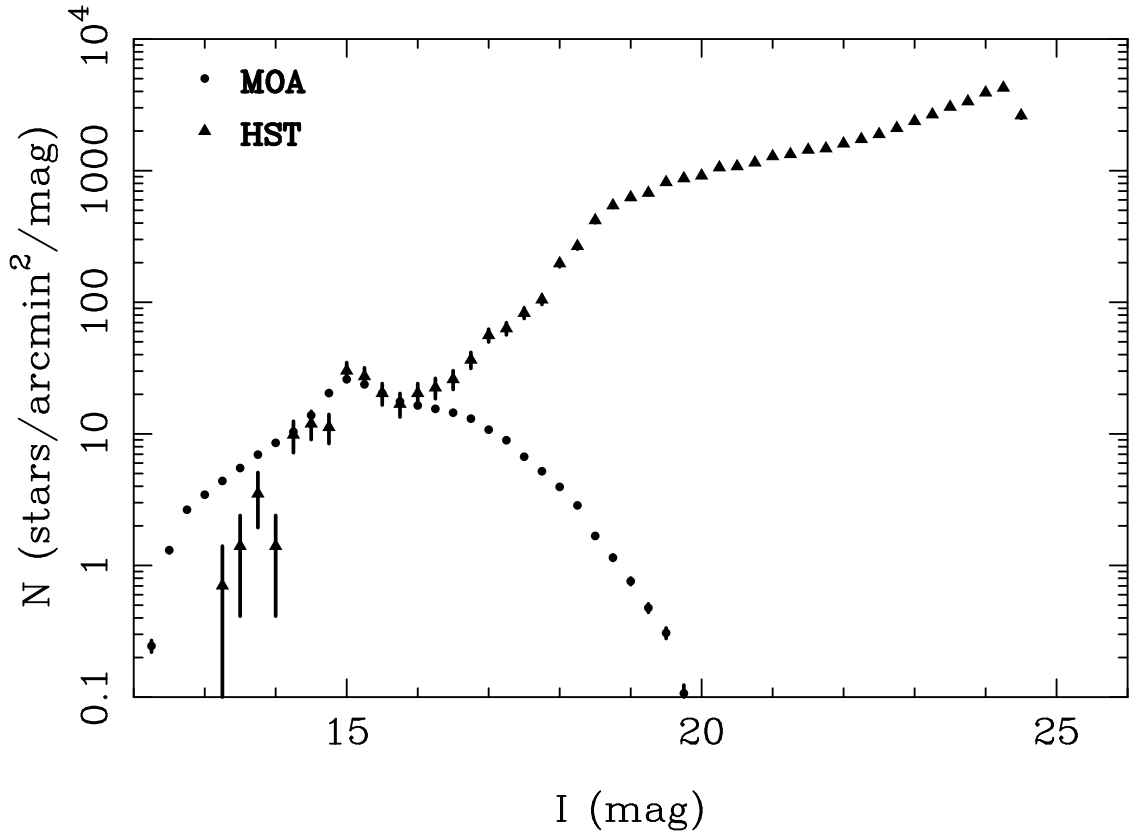}
\caption{Combined luminosity function of Baade's window from MOA
(filled circle) and HST (filled triangle).
For MOA data the offsets due to differential extinction are corrected
to match that of the HST field.
We use MOA data for $I<16$ and HST data for $I \ge 16$
in this analysis.}
\label{fig:lf}
\end{figure}

\begin{figure}
\includegraphics[angle=-90, scale=0.6]{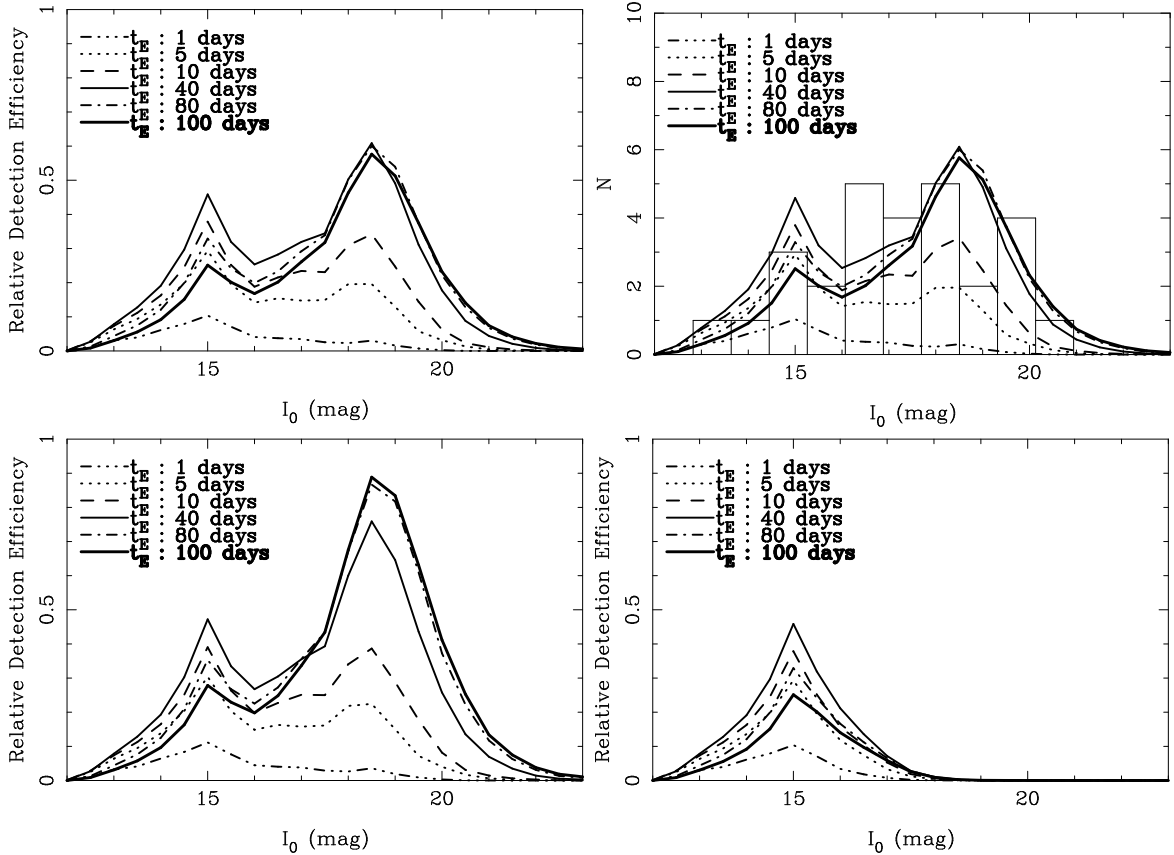}
\caption{Expected  relative event rates  as a function of the
source $I$-band magnitude  $I_0$ for the various event timescales as
indicated in Figure. Top-left: Expected event rates in this
analysis. Top-right: Histogram of the observed baseline
magnitude $I_0$ with corresponding expected event rates scaled to
match the histogram. The expected and observed distributions are
in good agreement for the observed range of $t_{\rm E}$ ($5<
t_{\rm E} <100$ days). Bottom-left: Expected event rates  without
the cut of $t_{\rm E} <200$ days in cut3. A significant increase
can be seen for the longer timescale events with the dimmer source
stars. Bottom-right: Expected event rates for the case that the
sources are only resolved stars in the reference image, which
corresponds to the event rate by DoPHOT-type analysis. This is
shown in comparison with that by DIA. }
\label{fig:eff_event_I}
\end{figure}

To emphasize this effect, we show the relative expected event rate
as a function of $I_0$ for various event
timescales in the top-left panel of  Figure \ref{fig:eff_event_I}.
These are estimated by multiplying Figure \ref{fig:eff_I} by  the
combined GB LF $\psi(I)$ (Figure \ref{fig:lf}), i.e., the 
integrand of equation (\ref{eq:eff}). The expected
source distributions are peaked  around $I\sim 19$ and go down to
$I\sim 22$, although the observational limiting magnitude of our
telescope and camera is $I\sim 17$. This does show the dramatic
capability of DIA. Especially note that this effect is increased
by the high sampling rate strategy that we have adopted. Though the
sensitivity of the MACHO group (\citealt{alc00}) became 2
magnitudes deeper than their observation limiting magnitude of
$V\sim 21$ by DIA, that of MOA is $\sim 5$ magnitude deeper than
our observation limiting magnitude of $I\sim 17$. (Of course, this
effect also depends on the shape of the luminosity function around
the observation limiting magnitude for each experiment.) The
longer the timescale is, the fainter the source events that could
be detected.

In the top-right panel in Figure \ref{fig:eff_event_I}, the observed 
$I$-band baseline magnitude $I_0$ distribution is shown as the histogram 
with corresponding expected event rates scaled to match the histogram.  
Here the observed $I_0$ is de-reddened to match that in the HST field.
The expected and observed distributions are in good agreement for the
observed $t_{\rm E}$ range ($5< t_{\rm E} <100$ days).

As mentioned in \S\,\ref{sec:cut3}, many possible microlensing
events are rejected by the condition $t_{\rm E} <200$ days in
cut3. These events have a significant increase in $\Delta$ F, but
their parameters could not be well constrained, because the light
curve have been measured only around the peak. If the baseline flux is
measured by the follow-up observations, $t_{\rm E}$ could be
constrained. If we make a catalogue of all variable stars and
DN as now we are doing, or if we observe with two colors, we
could easily identify the real microlensing events and issue
alerts without the  $t_{\rm E} <200$ days criterion.

We show the expected event rates without the cut $t_{\rm E} <200$
days in the bottom-left panel of Figure \ref{fig:eff_event_I}. In
this figure we can see a significant increase in the expected
event rate for the longer timescale events with the dimmer source
stars which  intrinsically should have high amplification. We can
see this effect in the minimum impact parameter distribution in
the following \S\,\ref{sec:eff_umin}. On the other hand, no
significant change can be seen for the short timescale events,
though the duration of the excess in the light curve of these
events are as short as that of the long timescale faint source
event. So, the efficiency for the short timescale events is not
affected by this criterion.

To compare these results with the traditional DoPHOT analysis in
which the source star should be resolved in the reference image,
we show the same distributions by using only the MOA luminosity
function (filled circle points in Figure \ref{fig:lf}) as the
source luminosity distribution in the bottom-right panel of Figure
\ref{fig:eff_event_I}. This is not actually the same case as with
DoPHOT, because the DIA can improve the photometric accuracy, even
for stars resolved in the reference image. Furthermore there are
many faint source events which are blended by bright resolved
stars. These effects will cancel each other. In any case they are
relatively small effects when compared with the differences in the
DIA and DoPHOT analysis. So this distribution should not be much
different from the actual one and make a meaningful comparison. In
this figure, the difference from DoPHOT (bottom-right) to DIA
(top-left) is clear. The peak of this distribution is at $I\sim
15$, which corresponds to the second peak in the top-left panel of
Figure \ref{fig:eff_event_I}. The DIA method increases the number
of events by more than a factor of 2. In particular we can detect
many events in which the source stars are fainter than the
observational limiting magnitude by using DIA.

\subsubsection{Minimum Impact Parameter Distribution.}
\label{sec:eff_umin}

Because of the simple geometry of microlensing, the probability
distribution of the minimum impact parameter $u_{min}$ is uniform.
In actual experiments, however, the distribution of the detection 
efficiency in $u_{\rm min}$, i.e. the expected event rate distribution, 
is biased towards smaller $u_{\rm min}$, because the high amplification 
makes it easier to detect the event.

\begin{figure}
\epsscale{0.8}
\includegraphics[angle=-90, scale=0.7]{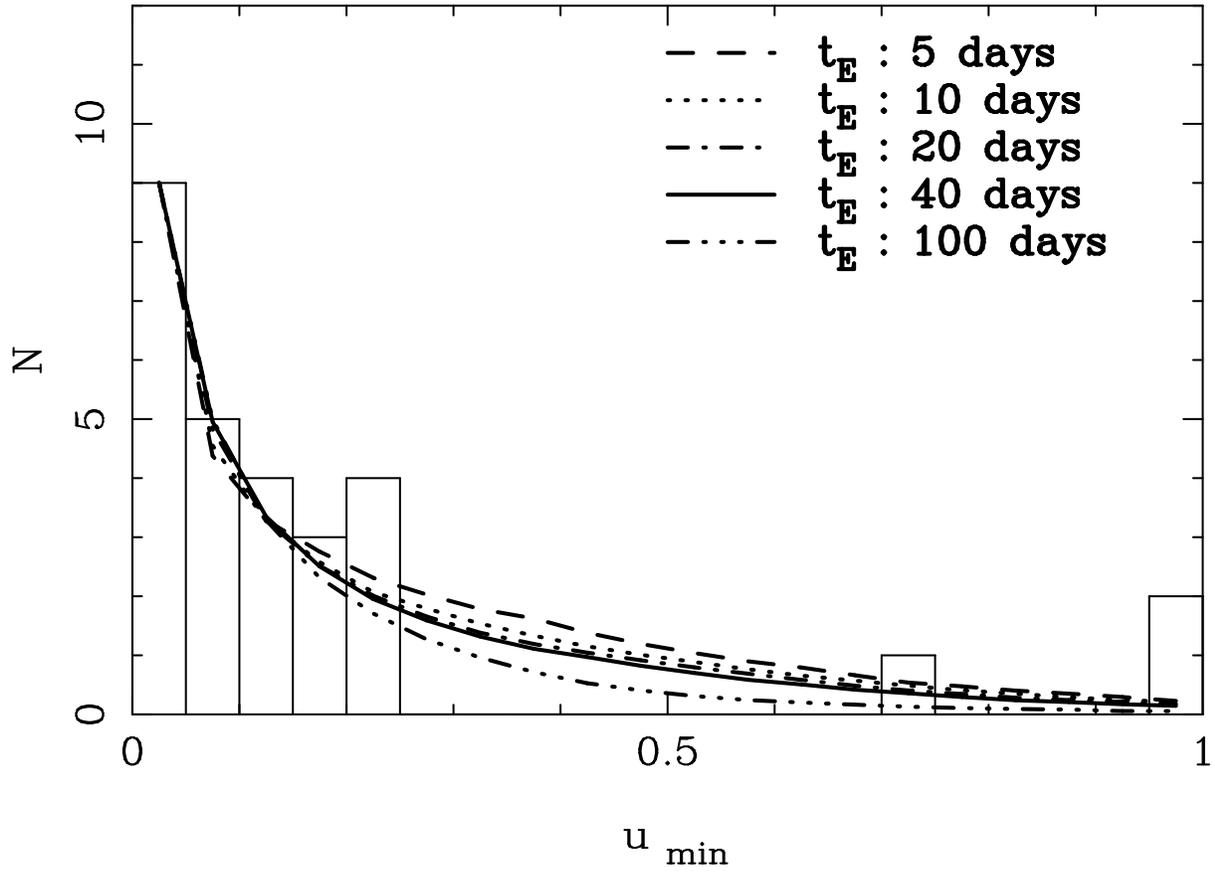}
\caption{Observed $u_{\rm min}$ distribution (histogram) and the
estimated distributions (lines) scaled to match the histogram at 
$u_{\rm min}=0.05$ for various timescales as indicated in the figure. }
\label{fig:umin_eff}
\end{figure}

This effect can be seen in our observed $u_{\rm min}$ distribution
(histogram) and the estimated one in the simulation (lines) for 
various timescales which are scaled to match the histogram at 
$u_{\rm min}=0.05$ as shown in Figure \ref{fig:umin_eff}.  Here the 
lines are the mean 
efficiencies for all our fields.  These observed and estimated 
distributions are in good agreement for the range of the observed 
timescales (between $5\sim 100$ days).

In this figure we can see that half of the detected events
have a small minimum impact parameter ($u_{\rm min} <0.1$), though
the fraction of events with $u<0.1$ is only $10\%$ in the DoPHOT analysis
(\citealt{alc97a}) and $30\%$ in  the recent DIA analysis
(\citealt{alc00}) by the MACHO group. This is because our sampling
rate is higher ($5\sim6$ times\,day$^{-1}$) than theirs (once per day).

\begin{figure}
\includegraphics[angle=-90, scale=0.7]{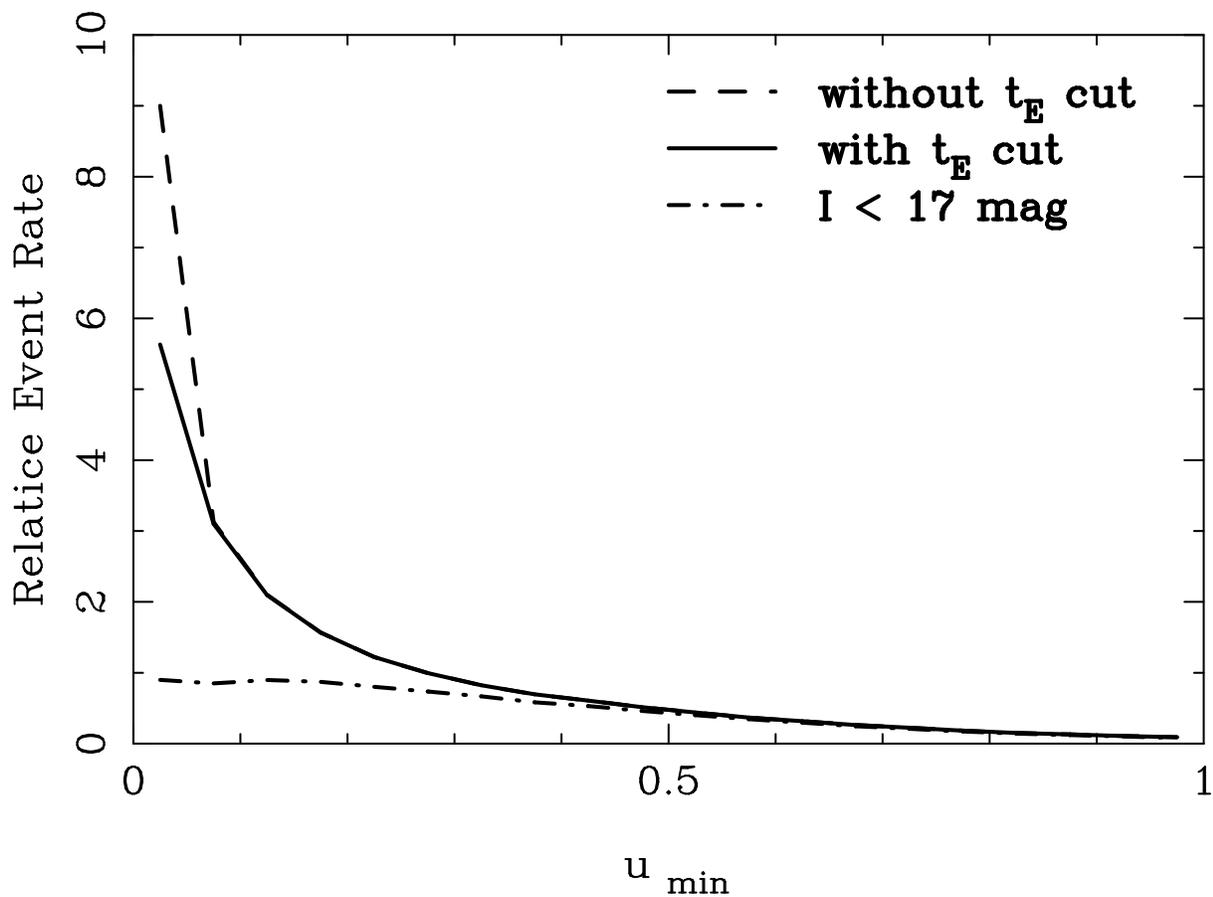}
\caption{Estimated
$u_{\rm min}$ distribution for the typical timescale of $t_{\rm E} = 40$ days
with (solid line)  and without (dashed line) the $t_{\rm E}<200$ days cut.
The same distribution with $t_{\rm E}<200$ days cut for the bright source
star ($I<17$) events (dot-dashed line), which is scaled to match the
others at $u_{\rm min}=1$, is also shown for comparison.}
\label{fig:umin_eff_various}
\end{figure}

If events which fail by the criterion $t_{\rm E} <200$ days, 
are included as mentioned in the previous section, then
$5\sim 15\%$ more high magnification events would be expected.
We show the estimated $u_{\rm min}$ distribution for the typical 
timescale of $t_{\rm E} = 40$ days without (solid line) and with 
(dashed line) this $t_{\rm E}<200$ days cut in Figure 
\ref{fig:umin_eff_various}. And we also show the same 
distribution with $t_{\rm E}<200$ days cut for brighter source 
stars ($I < 17$), which correspond to the DoPHOT analysis.

From this figure, it is clear that our analysis with DIA (solid line) can
detect the high magnification events more efficiently than a DoPHOT-type
analysis (dot-dashed line).
Furthermore it will be possible to detect more ($\sim 10\%$)
high magnification events without the $t_{\rm E}<200$ days cut (dashed line)
in the near future.

\subsection{Optical Depth}
\label{sec:Opticaldepth}

Here we estimate the optical depth towards the Galactic bulge by using
the observed events  and our detection efficiencies.

\subsubsection{Optical Depth Estimation}
\label{sec:opticaldepthestimation}

The optical depth $\tau$ is defined as the probability that any given
star is microlensed with impact parameter $u_{\rm min} \le 1$ at any given
time. The $\tau$ can be estimated by
\begin{equation}
  \label{eq:opt}
  \tau = \frac{\pi}{2N_{\rm s}T_{\rm o}} \sum_i \frac{t_{{\rm E},i}}{\varepsilon (t_{{\rm E},i})}
\end{equation}
where $N_{\rm s}$ is the total number of source stars in the observation 
fields and $T_{\rm o}$ is the duration of observation of this analysis in 
days. $t_{{\rm E},i}$ is the event timescale for the $i$th event and 
$\varepsilon (t_{{\rm E},i})$ is the detection efficiency for a given
timescale.

We estimate $N_{\rm s}$ for $I<23$ by using the number of bright stars
($I<15$ which is extinction corrected) whose observation efficiency is 
nearly $100\%$, and our combined LF.  This is because our detection 
efficiencies have been estimated by taking all stars with $I<23$ into 
account.

\begin{figure}
\includegraphics[angle=-90, scale=0.7]{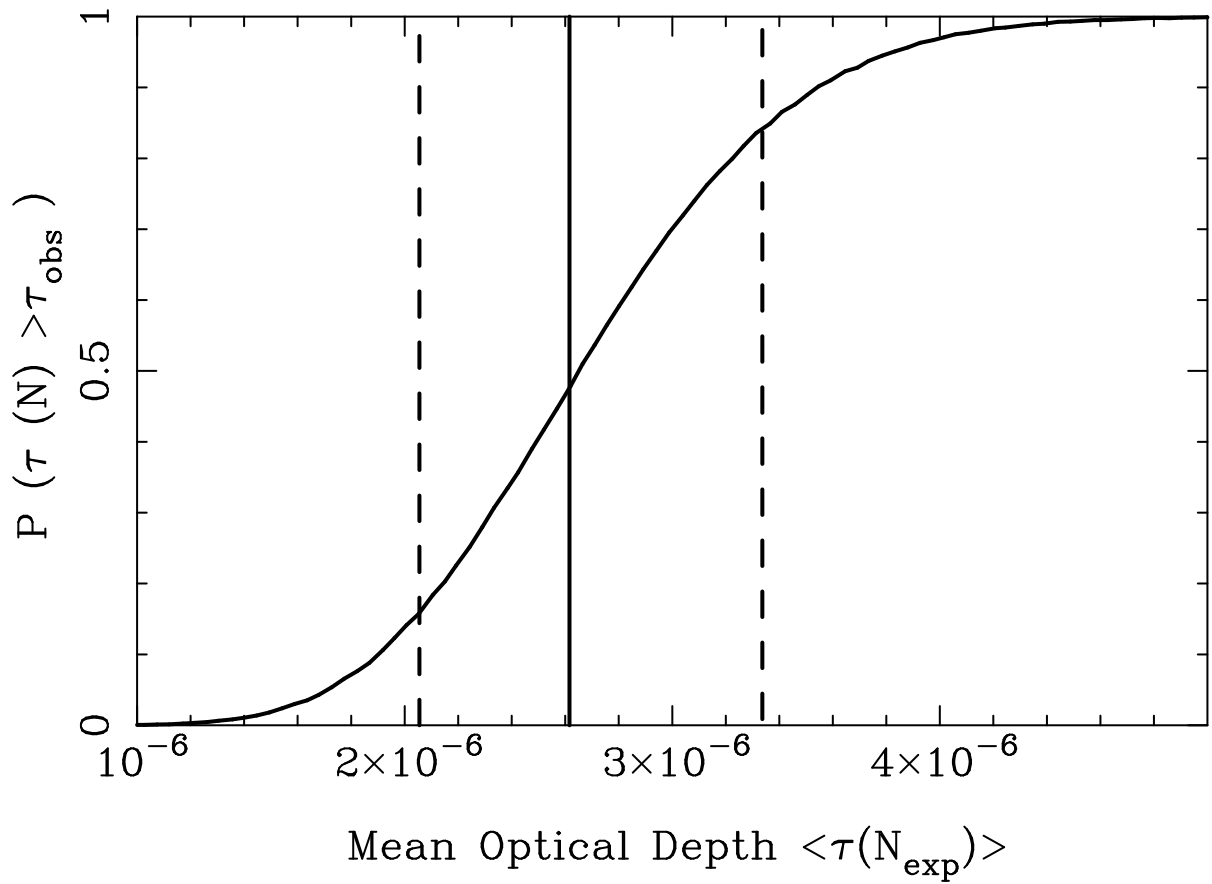}
\caption{
Probability $P(\tau(N) > \tau_{\rm obs})$ that the optical depth
$\tau(N)$ in each simulated experiment is larger than the observed one
$\tau_{\rm obs}$ as a function of the mean optical depth $\langle \tau(N_{\rm exp}) 
\rangle$ of experiments in which an expected event number is $N_{\rm exp}$.
$1\sigma$ confidence limits (dashed line) and the observed optical
depth $\tau_{\rm obs}$ (solid line) are also presented.}
\label{fig:opt_err}
\end{figure}

We estimate the optical depth by using the observed timescales
with  $N_{\rm s} \sim 250$ million stars ($I<23$), $T_{\rm o} =
219$ days for ngb1 $\sim$ 12. We do not include fields ngb13 and
14 in this analysis because these two fields are far from Baade's
Window. For events with timescales within $0.3<t_{\rm E}<200$
days, we estimate the optical depth as

\begin{equation}
\label{eq:tauobs}
\tau = 2.63_{-0.58}^{+0.72} \times 10^{-6},
\end{equation}
where the lower and upper limit of this value are estimated by a
Monte Carlo simulation following  \cite{alc97a,alc97b,alc00} and
assuming a Poisson distribution. 
We have simulated the 100,000 "experiments" for each $N_{\rm exp}$ 
which is the number of expected events and chosen at 0.5 intervals 
between 0 and 60 events.  In each experiment the number of detected 
events $N$ are selected by following the Poisson statistics with the 
mean value of $N_{exp}$.
For each simulated event, we randomly selected one of
our observed event timescales. We have estimated the probability
$P(\tau(N) > \tau_{obs})$ that the optical depth $\tau(N)$ in each
simulated experiment is larger than the observed one $\tau_{obs}$,
and the mean optical depth $\langle \tau(N_{exp}) \rangle$ for
each distribution with $N_{exp}$. We show this probability
distribution in Figure \ref{fig:opt_err}, from which we have estimated
the $1\sigma$ confidence limit of the optical depth.

\begin{figure}
\plotone{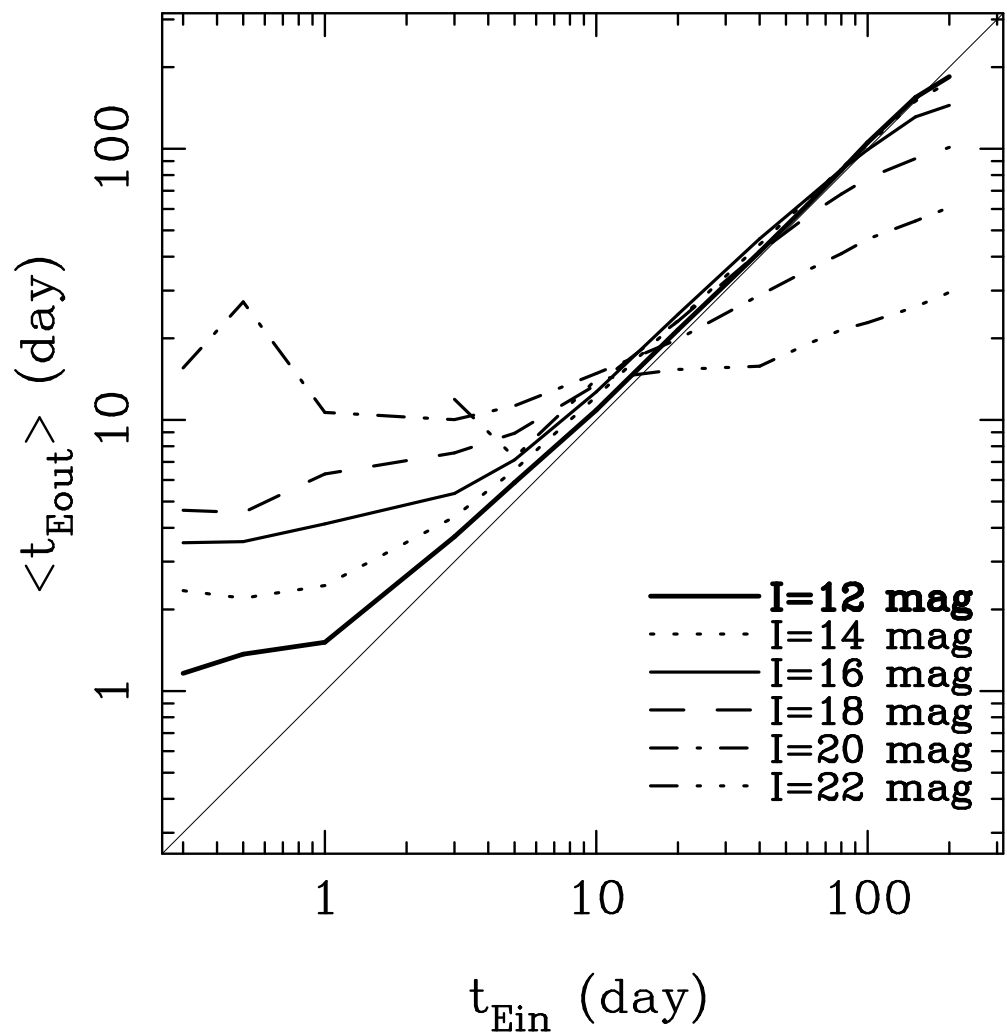}
\caption{
Relation between $t_{\rm Ein}$ and  the mean value of
$t_{\rm Eout}$ for various source I-band magnitude as indicated in the figure. 
The mean $<t_{\rm Eout}>$ are taken over the events between  
each indicated $I$ magnitude and $I+1$ mag. We have no detection 
in $t_{\rm Ein} \le 1$ day for $I=22$.
}
\label{fig:teinout}
\end{figure}

This estimate gives errors which are a lower limit, because errors in 
$t_{\rm E}$ are ignored. We found a large uncertainty and systematic bias
in the "INPUT" $t_{\rm E}$ (hereafter $t_{\rm Ein}$) and measured 
"OUTPUT" $t_{\rm E}$ (hereafter $t_{\rm Eout}$) in
our artificial events generated in \S\,\ref{sec:simulation}.  
We show the relation between $t_{\rm Ein}$ and  the mean value of 
$t_{\rm Eout}$ for various source magnitude in Fig. \ref{fig:teinout}.
As we can see in Fig. \ref{fig:teinout}, the measured $t_{\rm Eout}$
tend to be larger in small $t_{\rm Ein}$ and smaller in large $t_{\rm Ein}$.  
This is because we cut events with $t_{\rm Ein}$ outside $0.3<t_{\rm E}<200$ days.
This effect is larger especially for fainter source events because of 
their large scatter in $t_{\rm Eout}$. However, for relatively 
brighter source events with $10\le t_{\rm Ein}\le 60$ days, 
there is only a quite small bias that  $t_{\rm Eout}$ tend to be slightly larger.

\begin{figure}
\includegraphics[angle=-90, scale=0.7]{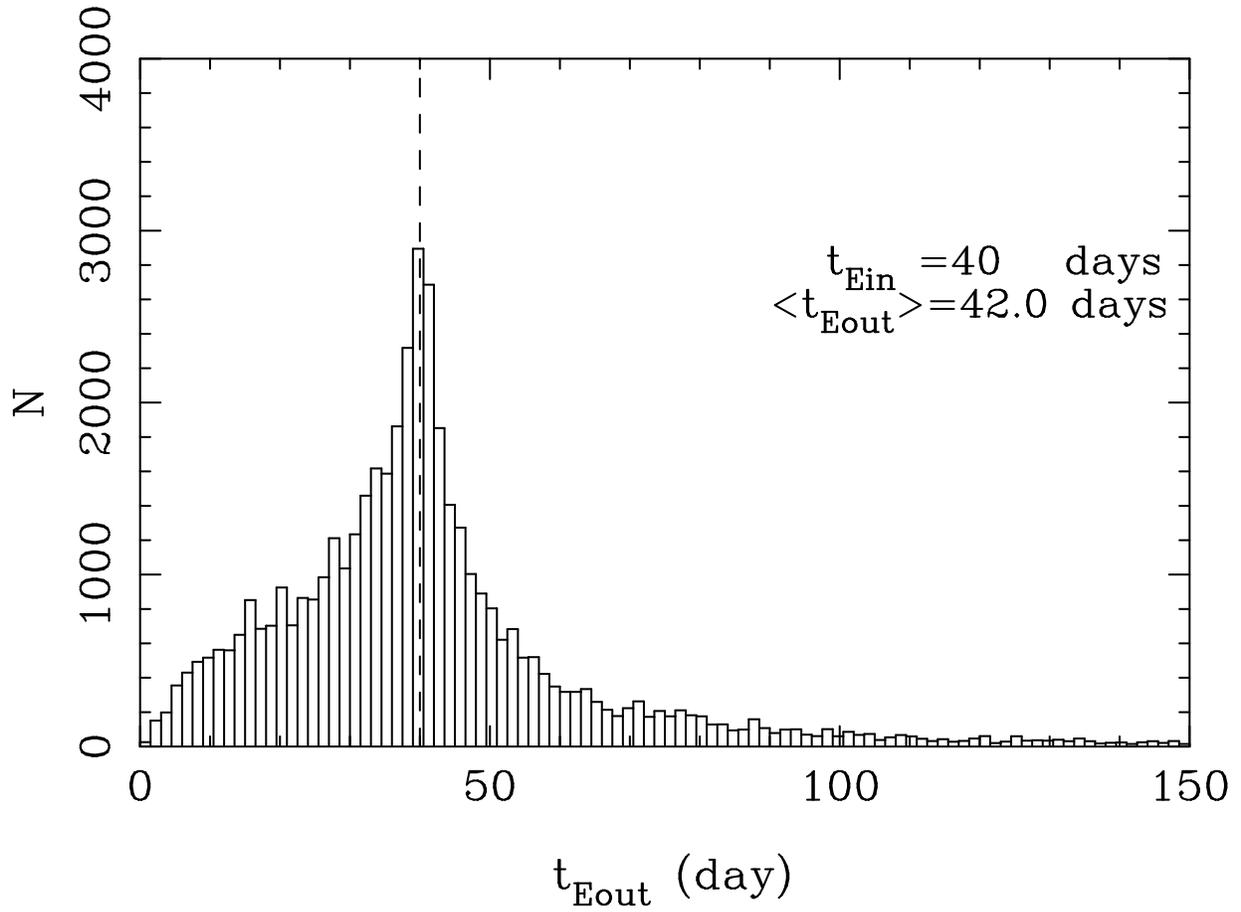}
\caption{ Example of the expected "OUTPUT" $t_{\rm Eout}$ distribution
for $t_{\rm Ein} = 40$ days (vertical dashed line), i.e.,
$D_{\rm out}(40)$. The mean $<t_{\rm Eout}>$ is 42 days in this case.
}
\label{fig:teout40}
\end{figure}

We carried out a full Monte-Carlo simulation to measure the
error in the optical depth and bias by using these artificial events.
We used lists of measured OUTPUT $t_{\rm Eout}$ of 
artificial events for each  "INPUT" value of $t_{\rm Ein}$ and $I_0$,
which consist of about a hundred thousand of artificial events for each
subfield.
The number of expected events for each $I_0$ is proportional to
$\sim \varepsilon_0(t_{\rm E}, I_0) \psi(I_0) u_{\rm th}$, i.e., the 
integrand of equation (\ref{eq:eff}).
Following this number of expected events for each $I_0$,
we picked up OUTPUT $t_{\rm Eout}$ randomly from the lists and put them 
into the expected OUTPUT $t_{\rm Eout}$ distribution 
$D_{\rm out}(t_{\rm Ein})$ until the total number of events 
become $\sim 40,000$ for each INPUT $t_{\rm Ein}$. 
We show a example of
this distribution for $t_{\rm Ein} = 40$ days, i.e. $D_{\rm out}(40)$,
in Fig. \ref{fig:teout40}.

We used our observed $t_{\rm E}$ distribution as the INPUT $t_{\rm Ein}$ 
distribution for the simulation. We don't have any artificial
events with exactly the same $t_{\rm E}$ as that of observed events. 
So we linked each event to artificial events which have a similar INPUT
$t_{\rm Ein}$ value.
For example, event ngb1-2-2745 ($t_{\rm E}=41.3$ day) was linked to 
artificial events with $t_{\rm Ein}=40$ day, and 
ngb1-2-5076 ($t_{\rm E}=51.8$ day) was linked to artificial events
with $t_{\rm Ein}=40$ and 60 days with 50\% probability.
The linked INPUT $t_{\rm Ein}$ of artificial events were listed in
the INPUT  $t_{\rm Ein}$ distribution $D_{\rm in}$.

We performed the same simulation as above by using $D_{\rm in}$ and 
$D_{\rm out}(t_{\rm Ein})$, i.e., we simulated  100,000 "experiments" 
for each $N_{\rm exp}$.
For each simulated event, we randomly select one of $t_{\rm Ein}$ 
from $D_{\rm in}$ and choose one of the $t_{\rm Eout}$ from corresponding 
OUTPUT distribution $D_{\rm out}(t_{\rm Ein})$ at random. Then we measure 
the INPUT and OUTPUT optical depths ($\tau_{\rm in}$ and $\tau_{\rm out}$) 
from $t_{\rm Ein}$ and $t_{\rm Eout}$ respectively, for each experiment.  
We ensured that the $\tau_{\rm in}$ in the case of $N_{\rm exp} = 28$ is 
the same as the observed value $\tau_{\rm obs}=2.63 \times 10^{-6}$.
This is expected because $D_{\rm in}$ was made from our observed
$t_{\rm E}$ distribution.

\begin{figure}
\includegraphics[angle=-90, scale=0.7]{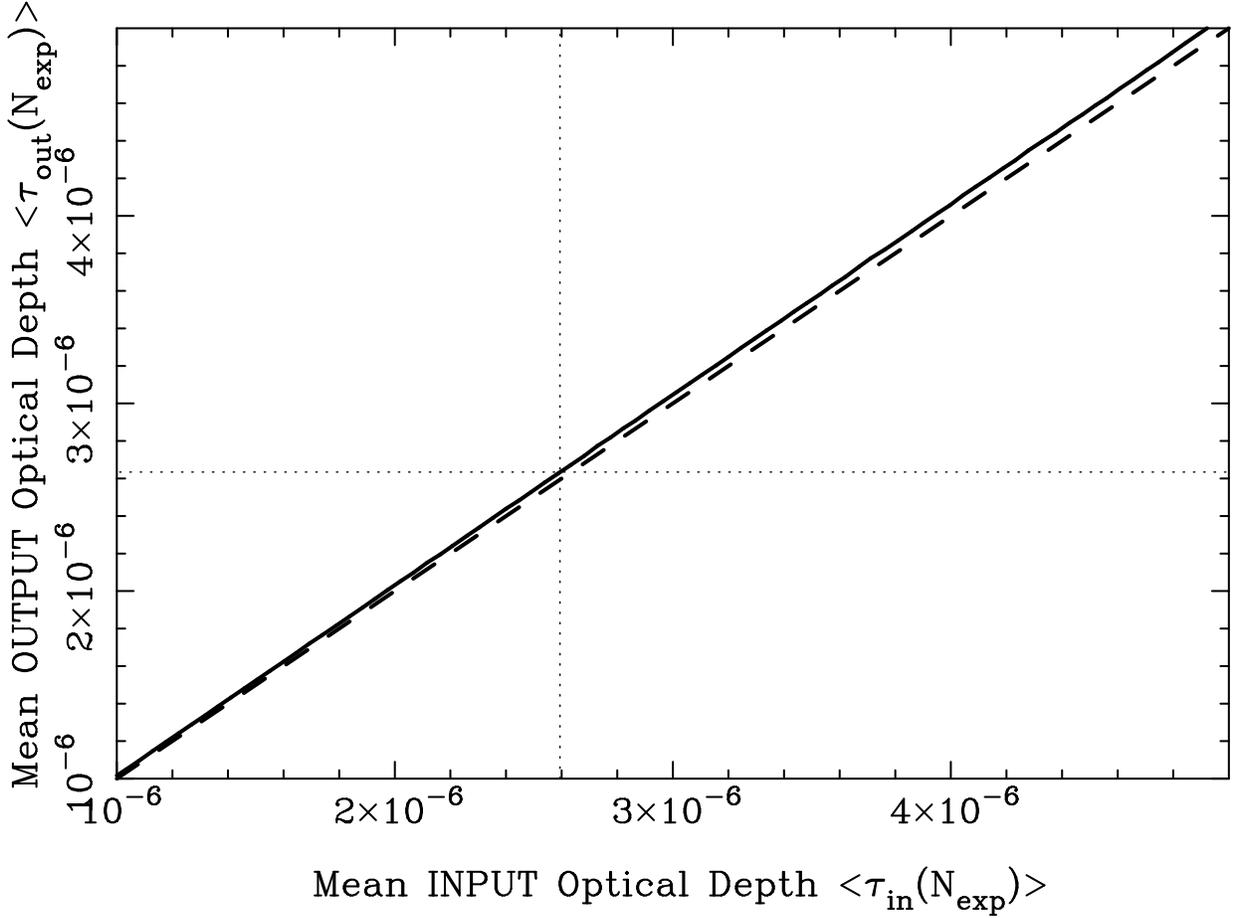}
\caption{ Relation between Mean optical depth of OUTPUT 
$<\tau_{\rm out}(N_{\rm exp})>$ and INPUT $<\tau_{\rm in}(N_{\rm exp})>$ 
(solid line). Dashed line represents 
$<\tau_{\rm out}(N_{\rm exp})> = <\tau_{\rm in}(N_{\rm exp})>$. 
The horizontal and vertical dotted 
line indicate observed ($\tau_{\rm obs} = 2.63\times 10^{-6}$) and 
estimated ($\tau = 2.59\times 10^{-6}$) optical depths respectively.
}
\label{fig:tauinout}
\end{figure}

We plot the difference between the mean value of $\tau_{\rm in}$ and 
$\tau_{\rm out}$ in Fig. \ref{fig:tauinout}.
We can see the existence of the small bias between them.
The $\tau_{\rm out}$ tend to be slightly larger than $\tau_{\rm in}$.
From this relation, the real optical depth is estimated to be
$\tau=2.59 \times 10^{-6}$ from the observed optical depth of 
$\tau_{\rm obs}=2.63 \times 10^{-6}$.

\begin{figure}
\includegraphics[angle=-90, scale=0.7]{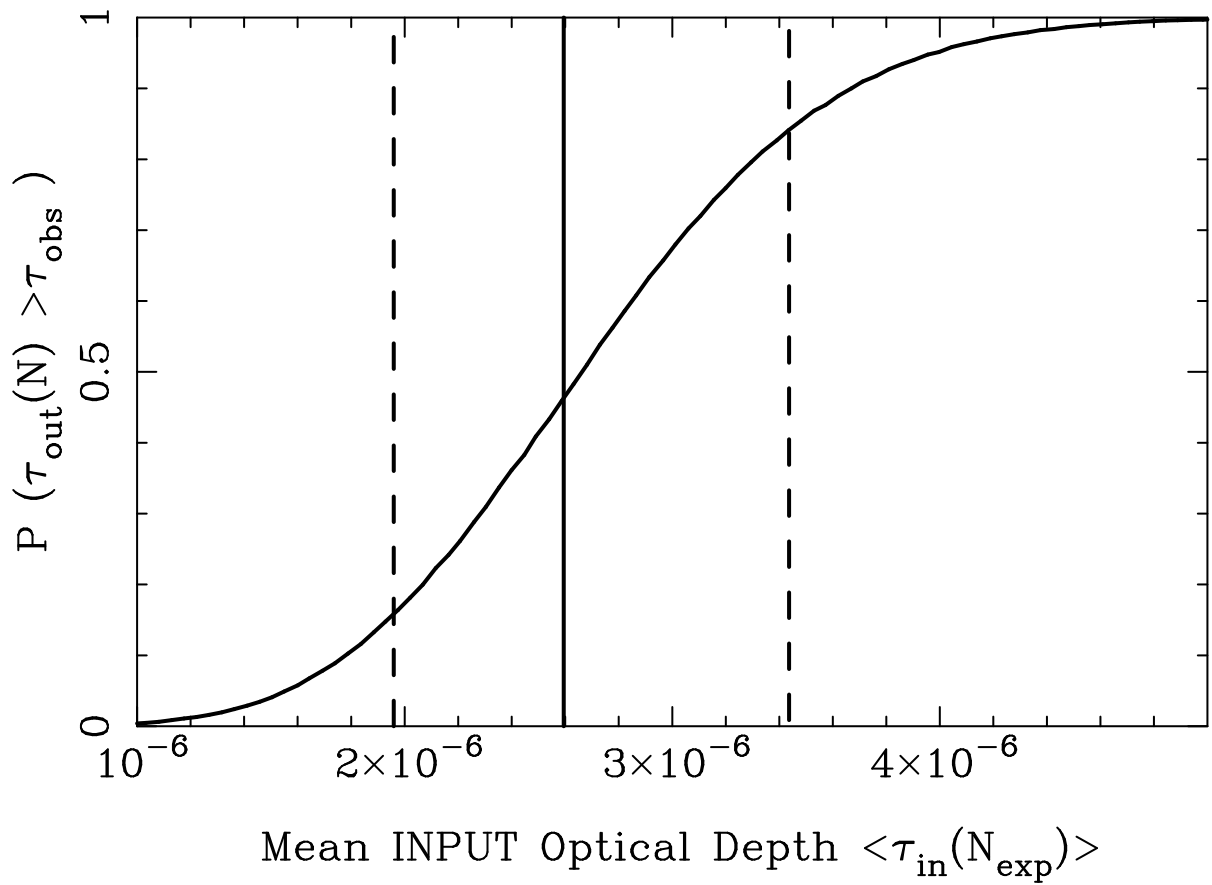}
\caption{ 
Probability $P(\tau_{\rm out}(N) > \tau_{\rm obs})$ that the OUTPUT optical depth
$\tau_{\rm out}(N)$ in each simulated experiment is larger than the observed one
$\tau_{\rm obs}$ as a function of the mean "INPUT" optical depth 
$\langle \tau_{\rm in}(N_{\rm exp}) \rangle$ of experiments in which an expected 
event number is $N_{\rm exp}$.
$1\sigma$ confidence limits (dashed line) and the estimated real optical
depth $\tau $ (solid line) are also presented.
}
\label{fig:opt_err2}
\end{figure}

To estimate the error in this optical depth, we have used the same method
as above but with $\tau_{\rm out}$ instead of $\tau_{\rm in}$.
The standard deviation of $\tau_{\rm out}$ is $\sim 16\%$ larger than that 
of $\tau_{\rm in}$ for each $N_{\rm exp}$.
We have estimated the probability $P(\tau_{\rm out}(N) > \tau_{\rm obs})$ 
that the optical depth $\tau_{\rm out}(N)$ in each simulated experiment 
is larger than the observed one $\tau_{\rm obs}$, and the mean INPUT optical 
depth $<\tau_{\rm in}(N_{\rm exp})>$ for each distribution with $N_{\rm exp}$. 
We show this probability distribution in Fig. \ref{fig:opt_err2}.
From this we have estimated the $1\sigma$ confidence limit of the 
"INPUT" optical depth $\tau_{\rm in}$.
These $1\sigma$ errors are  about 15\% larger than errors from
the method with Poisson statistics only.
The resultant optical depth and errors are

\begin{equation}
\label{eq:taunew}
\tau = 2.59_{-0.64}^{+0.84} \times 10^{-6}.
\end{equation}


This $\tau$ is mostly due to disk and bulge (bar) stars. The
optical depth in the direction of the Galactic bulge due to halo
objects of any kind is only $\sim 0.13 \times 10^{-6}$
(\citealt{gri91}). This $\tau$ is underestimated because some
fraction of the source stars are foreground disk stars, for which
the optical depth is considerably lower.

\subsubsection{Disk Contribution}
\label{sec:disk}

The optical depth estimated in the previous section is underestimated.
In the number of stars in our fields $N_s$, some fraction of the stars
are probably foreground Galactic disk stars. The optical depth for 
the disk source stars (so-called disk-disk events) is quite small.
There are also background disk stars which would have a higher optical 
depth, but our line of sight towards the bulge is several hundred parsecs
out of the Galactic plane on the far side of the bulge. So most of the 
disk contamination is from foreground disk stars. The fraction of disk 
stars in our fields is rather uncertain.

We estimate the fraction of disk stars out of all stars, 
$f_{\rm disk}$ in our fields. We use the non-rotating triaxial bar 
models with the bar inclination angle of $\theta =20^\circ$ whose 
density profile as a function of the distance $D$
from the Sun is given by (\citealt{han95a}; \citealt{alc00})

\begin{equation}
  \label{eq:bar}
  \rho_b(w) = \frac{M}{20.65abc}exp \left( -\frac{w^2}{2} \right),
\end{equation}

\begin{equation}
  \label{eq:bar_w}
  w^4 =\left[  \left(\frac{x'}{a}\right)^2 + \left(\frac{y'}{b}\right)^2  \right]^2 +
         \left(\frac{z}{c}\right)^4.
\end{equation}
Here the coordinates $(x',y')$ are measured along the longest axis and
another axis of the bar in the Galactic plane.
The $x'$-axis is aligned at an angle $\theta$ from the line of sight
to the Galactic center from the Sun, with the near side of the bar in
the positive-$l$ quadrant.
The $z$-axis is as usual the height above the Galactic plane.
These Galactocentric coordinates $(x',y',z)$ are given by
$x' = R_0 \cos\theta - D \cos b \cos(l+\theta)$,
$y' = R_0 \sin\theta - D \cos b \sin(l+\theta)$ and
$z  =  D \sin b$
where $R_0$ is the distance to the Galactic center from the Sun taken 
to $8.5$ kpc. $a=1580$ pc, $b=620$ pc and $c=430$ pc define the bar 
scale lengths, and $M=1.8\times10^{10} M_\odot$ is the total bar mass.

We use the standard double-exponential disk whose density profile is given by
(\citealt{han95a}; \citealt{alc00})

\begin{equation}
\label{eq:rhodisk}
\rho_d=\rho_{d0} exp\left( \frac{-|z|}{h_z}-\frac{R-R_0}{R_d} \right),
\end{equation}
where $R = R_0^2+D^2\cos^2b-2R_0D\cos b\cos l$ and $z= D \sin b$ are the disk
cylindrical Galactocentric coordinates, $D$ is the distance from the Sun
and $(l,b)$ are the Galactic latitude and longitude.
$h_z = 325$ kpc and $R_d = 3.5$ kpc are  the disk scale height and length,
$\rho_{d0} = 0.06M_\odot\,{\rm pc}^{-3}$ is the density constant chosen 
to match the density in the solar neighborhood.

These models give optical depths towards Baade's window 
$(l, b)=(1^\circ, -3^\circ.9)$ of $1.2\times 10^{-6}$ from the bar, 
and $0.6\times 10^{-6}$ from the disk.  By using these density models 
and \cite{kir94}'s luminosity function with $\beta=-1$
(see \S\,\ref{sec:timescale}), $f_{\rm disk}$ is estimated as $\sim 23\%$.
This value is consistent with the value used by the MACHO group of
$f_{disk} \sim 20\%$ (\citealt{alc97a}) and
$f_{disk} \sim 25\%$ (\citealt{alc00}), where
most of our GB fields are overlapping with their fields.
This gives

\begin{equation}
\label{eq:taubulge}
 \tau_{bulge} =  3.36_{-0.81}^{+1.11}\times 10^{-6}[0.77/(1-f_{\rm disk})].
\end{equation}
This optical depth is consistent with the previous observations of
$3.3^{+1.2}_{-1.2} \times 10^{-6}$ from $9$ events by DoPHOT analysis
(\citealt{uda94}), $3.9^{+1.8}_{-1.2} \times 10^{-6}$ from $13$ events
in the clump giant subsample by DoPHOT (\citealt{alc97a}) and
$3.23^{+0.52}_{-0.50} \times 10^{-6}$ from $99$ events
by DIA (\citealt{alc00}), while slightly higher than 
$2.0 ^{+0.4}_{-0.4} \times 10^{-6}$ (\citealt{pop00}) and
$2.23 ^{+0.38}_{-0.35} \times 10^{-6}$ (\citealt{pop02}).

This measured optical depth must be regarded as a lower limit of the
true value because our observations are only sensitive to the events
with $0.3< t_{\rm E} <200$ days. Nevertheless this value is still 
higher than those predicted by most Galactic models whose  mass and 
inclination of the bar are consistent with other observations.
We note that our observed optical depth values are averaged over 
$16$ deg$^2$ (12 fields) around Baade's Window 
$(l,b)=(1^\circ.0, -3^\circ.9)$, while in most calculations the optical
depth is estimated exactly towards Baade's Window. However we ignore 
this difference in this paper since it is negligible in most models.
Even the smallest inclination angle and a large bar mass have been 
reported to be insufficient to produce an optical depth greater than 
$\sim 2.5\times 10^{-6}$ (\citealt{pea98}).

\section{Timescale Distribution}
\label{sec:timescale}

\begin{figure}[htb]
\epsscale{0.7}
\includegraphics[angle=-90, scale=0.7]{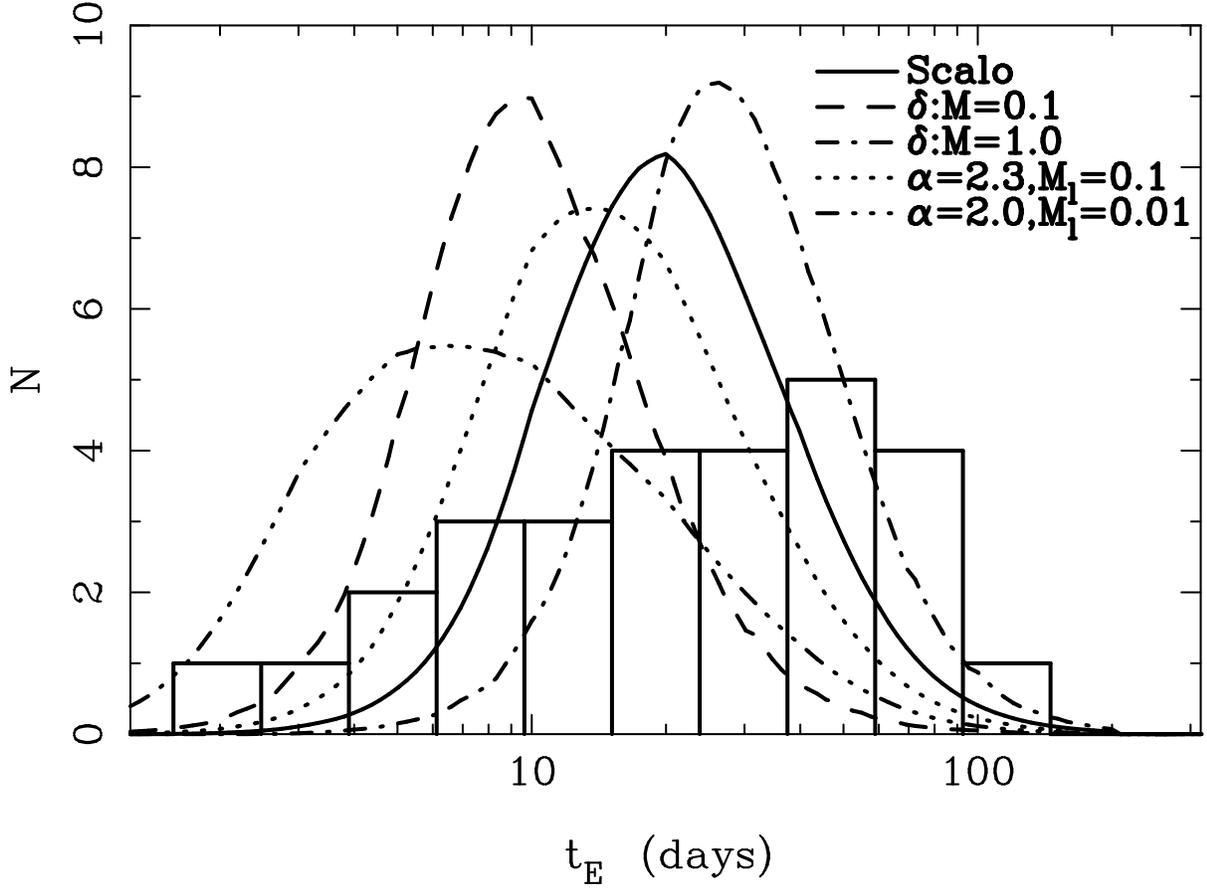}
\caption{Histograms of the event timescale $t_{\rm E}$
distribution for $28$ observed events with expected timescale
distributions, normalized to the observed number of events, for a
fixed bar and disk density model and for various mass functions.
The solid line indicates (i) \cite{sca86}'s PDMF; the dashed and
dot-dashed lines indicates $\delta$ function (ii) at
$M=0.1M_\odot$ and (iii) at $M=1.0M_\odot$ respectively; the
dotted and three dot-dashed lines show the power-law (iv) with
$\alpha=2.3$, $M_l = 0.1$ and (v) with $\alpha=2.0$, $M_l = 0.01$,
where the later one represents the brown dwarf rich mass function.}
\label{fig:plot_te}
\end{figure}

The observed timescale distribution depends on the mass function
and the velocity dispersion of the lens. To make meaningful models
of Galactic structure,  both the optical depth and the timescale
distribution should be consistent with the observations.
\cite{pea98} and \cite{mer98} show that it is difficult to
reproduce the observed timescale distribution of \cite{alc97a}
with the existing Galactic models. We show our observed timescale
distribution in Figure \ref{fig:plot_te}.

To be compared, we also plotted the expected timescale
distribution for a fixed bar (equation (\ref{eq:bar})) and disk
(equation (\ref{eq:rhodisk})) density model with various mass
functions in Figure \ref{fig:plot_te}. These distributions are
corrected by our detection efficiencies and normalized to the
number of observed events. Where we followed the method of  
\cite{kir94}, \cite{han95a} and\cite{alc00} assuming the mean velocity
$\bar{v}$ and variance $\sigma$ of the components in the bar and
disk for each direction ($y$ and $z$) as $(\bar{v}_{bar,y},
\bar{v}_{bar,z}) = (0,0)$ and $(\sigma_{bar,y},
\sigma_{bar,z})=(110,110)$ for the bar lens and source,
$(\bar{v}_{disk,y}, \bar{v}_{disk,z}) = (220,0)$ and
$(\sigma_{disk,y}, \sigma_{disk,z})=(30,30)$ for the disk lens,
$(\bar{v}_{o,y}, \bar{v}_{o,z}) = (220,0)$, $(\sigma_{o,y},
\sigma_{o,z})=(0,0)$ for the observer in km\,s$^{-1}$
(\citealt{han95a}; \citealt{alc00}). For these bar and disk
models, we evaluate timescale distributions towards Baade's
window with the following five mass functions: (i) \cite{sca86}'s
Present Day Mass Function (PDMF), a $\delta$-function (ii) at
$M=0.1M_\odot$ and (iii) at $M=1.0M_\odot$, and the power-law
$\phi(M) \propto M^{-\alpha}$ (iv) with $\alpha=2.3$ and the low
mass end $M_l = 0.1 M_\odot$, (v) with  $\alpha=2.0$ and $M_l =
0.01 M_\odot$, where the function (v) represents the brown dwarf
rich mass function (\citealt{alc97a}).

The timescale of detected events is distributed in the range 
$5<t_{\rm E} <100$ days and centered around $t_{\rm E} \sim 30$ 
days. This feature is consistent with that from the MACHO group 
(\citealt{alc00}) though their distribution is slightly sharper 
around the mean than ours.  The distribution of \cite{alc00} is 
well fitted by the timescale distribution expected from the 
\cite{sca86}'s PDMF except for some fraction of long timescale 
events, though those models do not explain the large observed 
optical depth. For our distribution in Figure \ref{fig:plot_te}, 
\cite{sca86}'s PDMF also seems to be more reasonable than the others.
Although our detection efficiency is sufficiently high for the
short timescale events ($t_{\rm E} \sim 0.3$ days) because of our 
frequent sampling ($5\sim 6$ times\,day$^{-1}$), there are only 
two short timescale events ($t_{\rm E} <4$ days).  The timescale 
distribution traces the mass distribution.  $t_{\rm E}$ for the 
most common events is $\sim 7 (M/0.1M_\odot)^{1/2}$ days for both 
bulge and disk lenses (\citealt{han95a}).  The number of observed 
short events is much smaller than that expected from the brown 
dwarf rich mass function (v), which is consistent with \cite{alc00}.
However, this topic is quite complicated. The measurement of 
$t_{\rm E}$ has large uncertainty in itself, and its distribution 
depends on  the unknown kinematics ofthe sources and lenses. 
A detailed analysis of this topic is beyond the scope of the present 
study. The identification of brown dwarfs may be possible only with
larger statistics of higher S/N events in the future.

We can also see some fraction of long timescale events ($t_{\rm E}
>70$ days) as reported in \cite{alc00}, which cannot be reproduced
with \cite{sca86}'s  mass function (i) and the contemporary
Galactic models (\citealt{han96}). Such long timescale events
could be produced when the lenses are heavy, moving at low
transverse velocity, or in the middle of the line from observer to
source. If both the lens and source are in the disk, so called
disk-disk lensing, then the timescale would be long. The
probability of this is constrained by star counts, and it is very
small.
There might be some unknown population of dynamically cold or
massive dark objects, such as white dwarfs or neutron stars in the
Galactic disk or bulge. More discussion on this can be seen in
\cite{alc00}.

\begin{figure}[htb]
\includegraphics[angle=-90, scale=0.7]{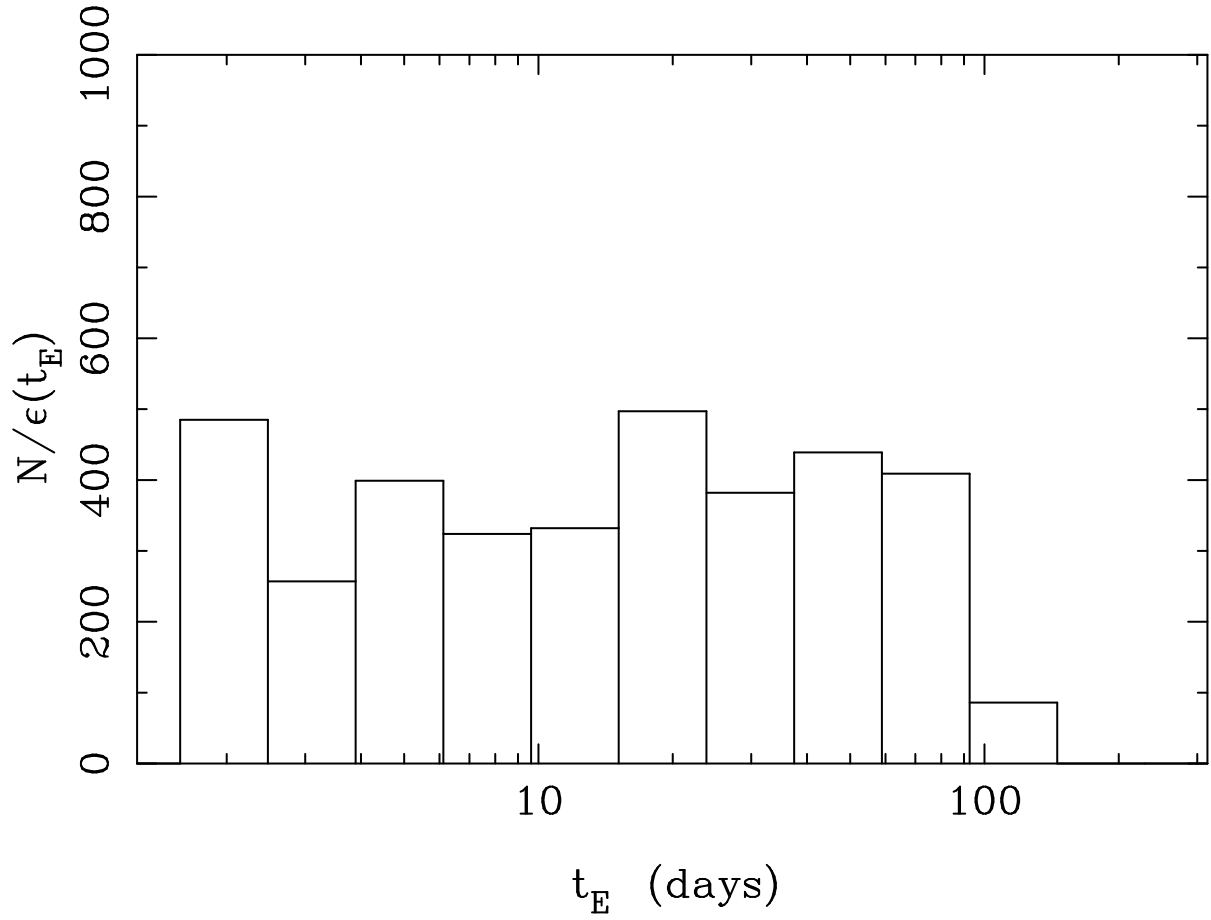}
\caption{Histograms of the observed $t_{\rm E}$ distribution,
corrected for the detection efficiency.}
\label{fig:plot_te_Neff}
\end{figure}

We show our observed timescale distribution, corrected by the 
detection efficiency, i.e., the expected true $t_{\rm E}$ 
distribution in Figure \ref{fig:plot_te_Neff}. This distribution
is similar to that in \cite{alc00} except the part for the short 
timescale events.  Though their distribution is sharply truncated 
at $t_{\rm E} \sim 4$ days except for one short event with 
$t_{\rm E} = 1.4$ days, our distribution is flat to $2$ days. 
However this difference is not significant because these shorter 
two bins are based on only two events, and the amount is still small.

\begin{figure}[htb]
\includegraphics[angle=-90, scale=0.7]{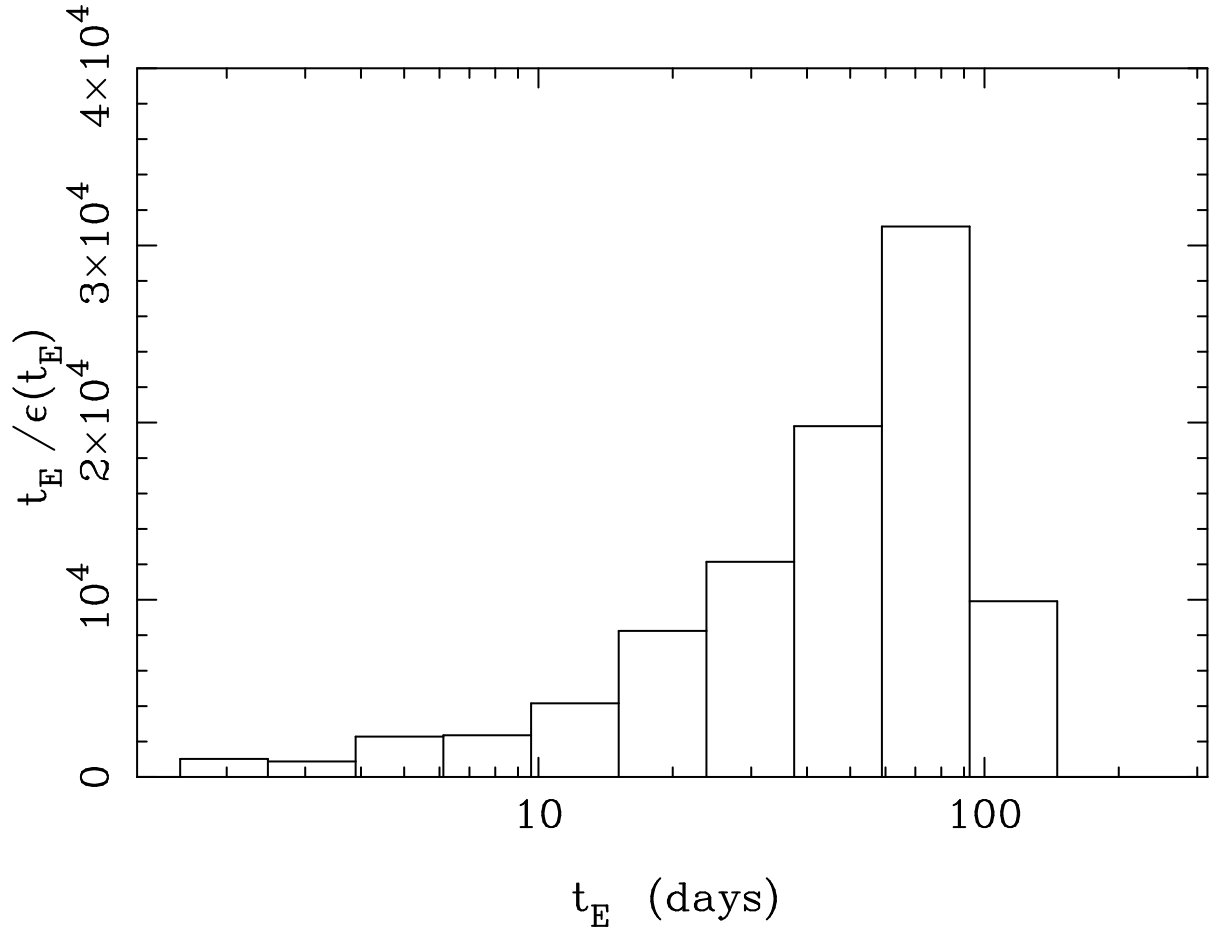}
\caption{Distribution of the contribution to the total optical depth
($t_{\rm E}/\epsilon(t_{\rm E})$) of the observed event $t_{\rm E}$
The contribution of the short timescale events to the
optical depth is  quite small. }
\label{fig:plot_te_eff}
\end{figure}

The contribution to the total optical depth of the observed
timescale distribution is given in Figure \ref{fig:plot_te_eff}.
The contributions of the short timescale events to the total
optical depth are quite small.

In any case, we need more observations to investigate the mass
function and Galactic structure in more detail.

\section{Discussion and Conclusion}
\label{sec:disc}

We have re-analyzed the sample of subtracted images that were derived
from the real-time DIA of GB observations obtained by MOA during 2000
(\cite{bon01}).  In this analysis we have found 28 microlensing event 
candidates in our 12 GB fields. The DIA is more suitable than DoPHOT 
analysis for our purpose, since the former method can detect the 
luminosity variation at any position, even where no star was previously
identified.

We have used a Monte Carlo simulation to estimate our event
detection efficiencies. By using these efficiencies and timescales
of our 28 detected events, we have estimated the optical depth towards
the GB for events with timescales within the range $0.3<t_{\rm
E}<200$ days as
$
\tau_{0.3}^{200} = 2.63_{-0.58}^{+0.72} \times 10^{-6},
$
where the statistical uncertainty of the optical depth have been 
estimated by Monte Carlo simulation only using the Poisson statistics.
This $\tau$ is overestimated and these errors in $\tau$ are underestimated.
We found the systematic bias and large uncertainty in the OUTPUT $t_{\rm Eout}$
in our simulation. By taking these bias and uncertainty in $t_{\rm Eout}$ into 
account, we get  

$$
\tau_{0.3}^{200} = 2.59_{-0.64}^{+0.84} \times 10^{-6}.
$$
The GB microlensing 
optical depth, in which the disk source stars component 
($f_{\rm disk}=23\%$) is taken into account, is given by

$$
 \tau_{\rm bulge} =  3.36_{-0.81}^{+1.11}\times 10^{-6}[0.77/(1-f_{\rm disk})].
$$
This value is  consistent with the previous observations of
$3.3_{+1.2}^{-1.2} \times 10^{-6}$ from $9$ events by DoPHOT analysis
(\citealt{uda94}), $3.9_{+1.2}^{-0.9} \times 10^{-6}$ from $13$ events 
in a clump giant subsample by DoPHOT (\citealt{alc97a}) and
$3.23_{+0.52}^{-0.50} \times 10^{-6}$ from $99$ events
by DIA (\citealt{alc00}), while slightly higher than  
$2.0^{+0.4}_{-0.4} \times 10^{-6}$ (\citealt{pop00}) and
$2.23 ^{+0.38}_{-0.35} \times 10^{-6}$ (\citealt{pop02}).

This observed $\tau$ must be regarded as  a lower limit of the
true value, since our observations are only sensitive to events with
$0.3< t_{\rm E} <200$ days. Nevertheless this value is still
higher than those predicted by most Galactic models whose  mass
and inclination of the bar are consistent with other observations.
Even the smallest inclination angle and a large bar mass could not
reproduce $\tau$ greater than $\sim 2.5\times 10^{-6}$
(\citealt{pea98}). In \cite{eva02} Freudenreich's model can reproduce 
the high optical depths toward the GB $\sim 2.5 \times 10^{-6}$ which 
is about 1 $\sigma$ level of our estimate.  

The smallest inclination produces the largest optical depth.
However the bar inclination has been reported to be in a wide
range between $10^\circ-45^\circ$ summarized in Table 7 in
\cite{alc00}. The optical depth also depends on the mass of the
bulge or bar. However various observations provide conflicting
values $M_{bulge|bar}=0.7-2.8 \times10^{10}$ $M_\odot$ (\cite{zha96};
\cite{dwe95}; \cite{hol98}; \cite{han95b}; \cite{blu95};
\cite{zha96}). Very small inclinations ($\theta=11^\circ$ with
$M_{bar}=2.0 \times10^{10}M_\odot$ (\citealt{zha96}) and
$\theta=12^\circ$ with $M_{bar}=2.5 \times10^{10}M_\odot$
(\cite{gyu99})) or a very heavy mass ($\sim 3.6\times
10^{10}M_\odot$ with $\theta = 20^\circ$ (\cite{gyu99})) are
required to account for the observed optical depth in this
analysis and in \cite{alc00}. \cite{bin01} estimate 
the minimum total mass in baryonic matter within the Solar circle to be
greater than $\sim 3.9\times 10^{10}M_\odot$ from $\tau = 2.0 \times 10^{-6}$,
and such high baryonic contribution is consistent with implications from
hydrodynamical modeling and the pattern speed of the Galactic bar.

Such a high mass would imply low halo MACHO fractions (\citealt{gat96}).
A massive bulge puts tight constraints on the contribution
of the disk to the rotation curve at small radii.
A small disk, however, leaves more room for the halo.
Since microlensing results towards the LMC fix the MACHO content in
the halo, a massive halo implies a smaller MACHO fraction.

The uncertainties in the Galactic bar orientation, the bar mass
and the stellar mass function are still large. So the optical
depth we have derived might yet be explained by other models. Further
discussion can be seen in \cite{alc00}.

We showed our observed timescale distribution which is not biased
by blending. 
Though our statistics are smaller, this seems to be consistent with 
the one previously presented by the MACHO group (\citealt{alc00}). 
The number of short timescale events was quite small, in spite of our 
high detection efficiency for events down to $t_{\rm E} \sim 0.3$ days.

A significant number of long timescale ($t_{\rm E} >70$ days)
events have been detected, as reported previously by the MACHO group
(\citealt{alc00}). These could not be explained by any current
Galactic model (\citealt{han96}). These might be a heavier remnant
component, such as white dwarfs, or some dynamically cold component. 
Either way, we need more observations to investigate the mass
function and Galactic structure in greater detail.

We have shown how efficiently MOA can detect the high magnification
events in which the probability of detecting extrasolar planets is
high and find $50\sim 60 \%$  of all detected events have high
magnification ($u_{\rm min} <0.1$). This fraction is much higher than
$10\%$ from the DoPHOT analysis (\citealt{alc97a}) and $30\%$ from
recent DIA analysis (\citealt{alc00}) by the MACHO group. This is
because our sampling  rate is higher ($5\sim6$ times\,day$^{-1}$)
than theirs. These results support our belief that high frequency
observations and analysis using DIA, that MOA is currently
carrying out, can detect high magnification microlensing events
very efficiently, even with a small telescope.

\vspace{5cm}

\acknowledgments 
We acknowledge J.~Holtzman, K.~Stanek, the MACHO
collaboration and the OGLE collaboration for making their data
publicly available. 
We would like to thank B.~Paczy\'{n}ski, S.~Mao and 
M. Smith for carefully reading the manuscript and comments.
We acknowledge A.~Drake for helpful comments. 
We are also thankful to the referee for suggestive comments.
Financial support of the Royal Society of New Zealand's Marsden Fund,
Grant-in-Aid for Scientific Research A, B and data base of Japanese 
Ministry of Education and Science are gratefully acknowledged.
T.S. acknowledge the financial support from the Nishina Memorial Foundation.

\begin{figure}
\plotone{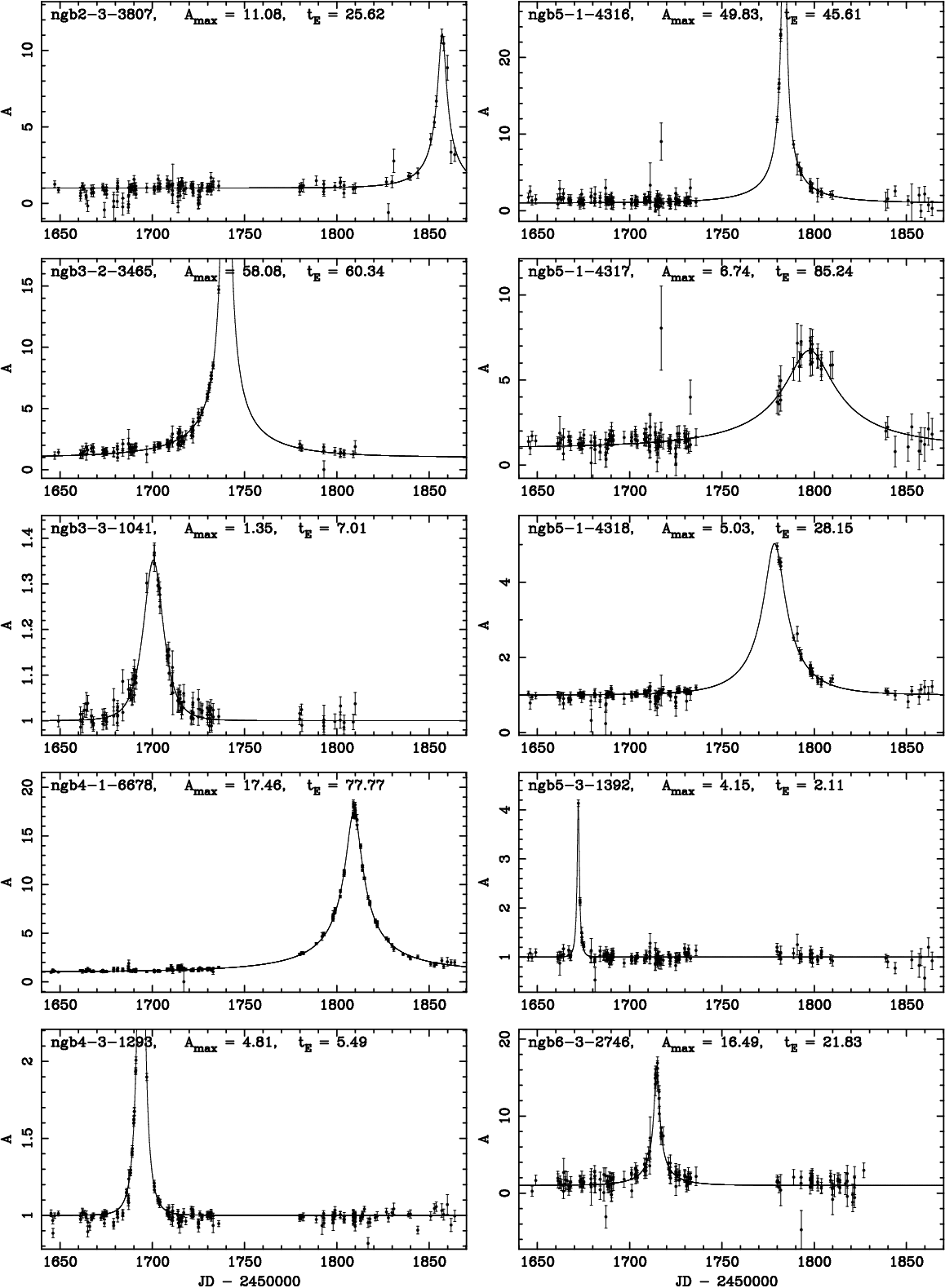}
\caption{Light curves of microlensing event candidates.}
\label{fig:lc11-20}
\end{figure}

\clearpage

\begin{deluxetable}{cc}
\tablecaption{Selection criteria in cut1. \label{tbl:cut1}}
\tablewidth{0pt}
\tablehead{
   \colhead{type}       &  \colhead{criteria} }
\startdata
      type 1     & $N_{det,n} = 0$ \\
                 & $N_{det,p}\ge 2$ \\
                 & $N_{clus,p} < 4$ \\
                 & $N_{det,p} \ne N_{clus,p}$ \\
                 & $Ratio_1 \le 0.1$ or $Ratio_3 \le 0.1$ \\
    \hline
      type 2     & $N_{det,p} = 0$ \\
                 & $N_{det,n}  \ge N_{frame}/4$  \\
                 & $N_{clus,n} < 5$ \\
    \hline
      type 3     & $N_{det,n} \ne 0$ and $N_{det,p} \ne 0$ \\
                 & $N_{clus,p} < 4$ \\
                 & $N_{clus,n} < 5$ \\
                 & ($N_{det,p}\ge 2$ and $N_{det,p} \ne N_{clus,p}$)
                                    or $N_{det,n}  \ge N_{frame}/4$ \\
\enddata
\end{deluxetable}

\begin{deluxetable}{c}
 \tablecaption{Selection criteria in cut2.  \label{tbl:cut2}}
\tablewidth{0pt}
\tablehead{    \colhead{criteria}  }
\startdata
            $N_{total} \geq 70$ \\
            $N_{in} >3$ \\
            $N_{out} > 9$ \\
            $1 \le N_{peak} \le 3$ \\
            $\sum_{i,peak} \sigma_i \ge 20$ \\
            $N_{hi} < 6$ \\
            $\chi^2_{out} <4$ or $\chi^2_{in}/\chi^2_{out} \ge 15$   \\
\enddata
\end{deluxetable}

\begin{deluxetable}{c}
\tablecaption{Selection criteria in cut3.  \label{tbl:cut3}}
\tablewidth{0pt}
\tablehead{
\colhead{criteria}
}
\startdata
            $\chi^2/d.o.f. < 3.5$  for $F_{\rm peak} < 450,000$ ADU \\
            $\chi^2/d.o.f. < 100$  for $F_{\rm peak} > 450,000$ ADU \\
            $u_{\rm min} < 1$ \\
            $0.3<t_{\rm E} < 200$ day \\
\enddata
\end{deluxetable}

\begin{deluxetable}{@{}c@{}c@{}ccccc}
\tablecaption{Position of 28 microlensing event candidates
    \label{tbl:candidates}}
\tablewidth{0pt}
\tablehead{
\colhead{field} & \colhead{chip} & \colhead{ID} & \colhead{ID (real-time)} &
      \colhead{ID (alert)} & \colhead{RA (J2000)} & \colhead{Dec (J2000)}
}
\startdata
    ngb1  & 2 & 2745 & -    & -           & 17:58:13.136 & -29:09:14.23 \\
    ngb1  & 2 & 4925 & 2717 & 2000-BLG-11 & 17:57:07.907 & -29:09:59.28 \\
    ngb1  & 2 & 5076 & -    & -           & 17:56:33.952 & -29:27:16.08 \\
    ngb1  & 2 & 5157 & 2667 & 2000-BLG-7  & 17:54:56.681 & -29:31:47.50 \\
    ngb1  & 3 & 2567 &  727 & 2000-BLG-3  & 17:54:29.770 & -28:55:59.31 \\
    ngb1  & 3 & 6328 & 2540 & -           & 17:58:20.936 & -28:47:48.76 \\
    ngb1  & 3 & 6344 & 2548 & -           & 17:55:05.425 & -28:50:34.60 \\
    ngb1  & 3 & 7416 & -    & -           & 17:57:54.728 & -28:54:32.66 \\
    ngb2  & 2 & 3867 & 1648 & -           & 18:00:12.361 & -29:37:23.93 \\
    ngb2  & 3 & 1932 & -    & -           & 17:59:00.087 & -29:33:01.11 \\
    ngb2  & 3 & 3807 & -    & -           & 18:00:07.092 & -29:23:27.47 \\
    ngb3  & 2 & 3465 & 1316 & 2000-BLG-9  & 18:05:09.533 & -30:36:06.77 \\
    ngb3  & 3 & 1041 & -    & -           & 18:06:47.640 & -29:50:09.57 \\
    ngb4  & 1 & 6678 & 2806 & 2000-BLG-13 & 17:55:33.202 & -28:10:17.09 \\
    ngb4  & 3 & 1293 &  159 & -           & 17:57:47.197 & -27:33:52.80 \\
    ngb5  & 1 & 4316 & 1673 & -           & 18:01:06.711 & -28:52:22.28 \\
    ngb5  & 1 & 4317 & 1672 & -           & 18:01:26.814 & -28:52:34.66 \\
    ngb5  & 1 & 4318 & 1668 & -           & 18:01:44.791 & -28:58:03.53 \\
    ngb5  & 3 & 1392 & -    & -           & 18:01:21.393 & -28:02:51.70 \\
    ngb6  & 3 & 2746 & -    & -           & 18:04:46.138 & -28:31:31.54 \\
    ngb6  & 3 & 3954 & 1425 & 2000-BLG-12 & 18:03:54.776 & -28:34:58.62 \\
    ngb7  & 3 & 2192 &  703 & 2000-BLG-8  & 18:10:55.621 & -29:03:54.20 \\
    ngb9  & 3 & 2336 &  841 & -           & 18:10:17.990 & -27:31:19.31 \\
    ngb10 & 1 & 1837 & -    & -           & 18:08:32.453 & -26:09:29.66 \\
    ngb10 & 3 & 2112 & -    & -           & 18:08:51.926 & -25:24:40.46 \\
    ngb11 & 2 & 1594 & 1142 & 2000-BLG-10 & 18:11:28.310 & -26:15:05.81 \\
    ngb11 & 3 & 1063 & -    & -           & 18:11:57.020 & -25:54:57.32 \\
    ngb12 & 2 & 3187 & 1052 & -           & 18:14:47.421 & -25:32:53.64 \\
\enddata
\tablecomments{ID in this offline analysis, ID in real-time analysis and
    alert ID in \cite{bon01} are also presented.
}
\end{deluxetable}

\begin{deluxetable}{ccccccccccc}
\scriptsize
\tablecaption{
Parameters in microlensing light curve fitting for 28 candidates
    \label{tbl:parameters}}
\tablewidth{0pt}
\tablehead{
   \multicolumn{3}{c}{MOA ID} & \multicolumn{1}{c}{$t_{0}$} &
   \multicolumn{3}{c}{$A_{max}$} & \multicolumn{3}{c}{$t_{\rm E}$ (day)} & $I_0$ \\
 field & chip & ID & JD-245000 & lower & best & upper & lower & best & upper & mag
}
\startdata
ngb1 & 2 & 2745 & 1680.16 & 200.0 & $2.6\times10^6$ & $8.8\times10^8$ & 26.9 & 41.3 & - & 19.7 \\
ngb1 & 2 & 4925 & 1799.38 & 8.5 & 8.6 & 8.7 & 56.9 & 57.2 & 57.5 & 13.6 \\
ngb1 & 2 & 5076 & 1725.17 & 56.1 & 91.7 & 234.4 & 33.2 & 51.8 & 106.4 & 20.8 \\
ngb1 & 2 & 5157 & 1725.79 & 4.1 & 5.1 & 6.2 & 2.9 & 3.4 & 4.0 & 17.1 \\
ngb1 & 3 & 2567 & 1691.82 & 2.9 & 9.4 & 293.9 & 6.9 & 16.0 & - & 17.5 \\
ngb1 & 3 & 6328 & 1795.55 & 9.4 & 12.5 & 16.8 & 30.6 & 37.2 & 45.9 & 17.6 \\
ngb1 & 3 & 6344 & 1792.80 & 14.9 & 28.4 & 60.3 & 10.0 & 15.4 & 27.3 & 18.3 \\
ngb1 & 3 & 7416 & 1829.40 & 10.6 & 11.9 & 13.2 & 10.6 & 11.4 & 12.2 & 16.2 \\
ngb2 & 2 & 3867 & 1801.09 & 4.6 & 6.1 & 9.1 & 92.0 & 116.2 & 163.6 & 17.4 \\
ngb2 & 3 & 1932 & 1707.91 & 76.5 & 140.6 & 361.4 & 21.1 & 36.3 & 181.2 & 19.9 \\ngb2 & 3 & 3807 & 1857.06 & 7.2 & 11.1 & 17.6 & 18.1 & 25.6 & 37.9 & 17.8 \\
ngb3 & 2 & 3465 & 1739.96 & 28.7 & 58.1 & 115.9 & 45.9 & 60.3 & 269.2 & 18.8 \\
ngb3 & 3 & 1041 & 1700.57 & 1.2 & 1.4 & 1.9 & 0.3 & 7.0 & 9.6 & 15.1 \\
ngb4 & 1 & 6678 & 1809.17 & 16.4 & 17.5 & 18.6 & 73.4 & 77.8 & 82.4 & 16.2 \\
ngb4 & 3 & 1293 & 1693.95 & 3.0 & 4.8 & 12.8 & 4.8 & 5.5 & 6.2 & 14.5 \\
ngb5 & 1 & 4316 & 1783.60 & 33.5 & 49.8 & 89.0 & 35.2 & 45.6 & 62.8 & 18.1 \\
ngb5 & 1 & 4317 & 1797.46 & 3.8 & 6.7 & 16.9 & 55.2 & 85.2 & 188.2 & 17.9 \\
ngb5 & 1 & 4318 & 1778.51 & 3.3 & 5.0 & 6.8 & 22.0 & 28.2 & 34.2 & 16.0 \\
ngb5 & 3 & 1392 & 1672.15 & 2.4 & 4.1 & 5.8 & 1.6 & 2.1 & 3.0 & 16.5 \\
ngb6 & 3 & 2746 & 1714.69 & 11.5 & 16.5 & 24.6 & 16.3 & 21.8 & 30.9 & 19.7 \\
ngb6 & 3 & 3954 & 1796.00 & 6.5 & 7.5 & 8.8 & 12.4 & 13.4 & 14.5 & 16.5 \\
ngb7 & 3 & 2192 & 1732.84 & 1.5 & 1.7 & 1.8 & 12.7 & 13.5 & 14.4 & 14.6 \\
ngb9 & 3 & 2336 & 1792.14 & 1.2 & 1.4 & 1.9 & 0.8 & 7.5 & 10.3 & 14.1 \\
ngb10 & 1 & 1837 & 1663.93 & 38.2 & 51.8 & 71.5 & 12.7 & 15.4 & 19.0 & 19.0 \\
ngb10 & 3 & 2112 & 1700.85 & 15.0 & 28.8 & - & 2.9 & 5.8 & 99.4 & 18.3 \\
ngb11 & 2 & 1594 & 1730.46 & 9.0 & 9.6 & 10.4 & 45.7 & 48.8 & 52.3 & 16.6 \\
ngb11 & 3 & 1063 & 1685.29 & 16.5 & 27.2 & 82.4 & 25.6 & 38.4 & 107.1 & 19.7 \\
ngb12 & 2 & 3187 & 1789.45 & 3.7 & 4.6 & 5.6 & 6.7 & 7.5 & 8.2 & 15.8 \\
\enddata
\tablecomments{$I$-band baseline magnitude of source star $I_0$ is
     de-reddened to match the HST field by using the $I$-band extinction
     $A_I$ map of each field (see \S\,\ref{sec:extinction}).
     The symbol ``-'' means that a value could not be constrained.
}
\end{deluxetable}

\end{document}